\documentclass[acmlarge, authorversion]{acmart}

\setcopyright{acmlicensed}
\copyrightyear{2025}
\acmYear{2025}
\acmDOI{XXXXXXX.XXXXXXX}

\acmJournal{TIOT}
\acmVolume{6}
\acmNumber{2}
\acmArticle{1}
\acmMonth{3}

\usepackage[inline]{enumitem}
\usepackage[nolist]{acronym}
\usepackage[capitalise]{cleveref}

\Crefname{equation}{Eq.}{Eqs.}
\Crefname{figure}{Fig.}{Figs.}
\Crefname{tabular}{Tab.}{Tabs.}

\def\system{automotive software system}

\def\simulators{simulators}

\usepackage[most]{tcolorbox}

\newtcbtheorem[auto counter]{finding}{Finding}%
{
    enhanced,
    colback=black!5,
    frame hidden,
    boxrule=0pt,
    borderline west={2pt}{0pt}{white!20!black},
    sharp corners,
    detach title,
    before upper=\tcbtitle \\ \par,
    coltitle=black!85,
    fonttitle=\bfseries,
    description font=\mdseries,
}{t!}

\usepackage{textpos}
\begin{document}
\begin{textblock}{12}(0,12)
    \centering \scriptsize
    This work has been submitted to the Association for Computing Machinery (ACM) for possible publication. Copyright may be transferred without notice, after which this version may no longer be accessible.\hfill
\end{textblock}

%%
%% The "title" command has an optional parameter,
%% allowing the author to define a "short title" to be used in page headers.
\title{Simulation to Reality: Testbeds and Architectures for Connected and Automated Vehicles}

%%
%% The "author" command and its associated commands are used to define
%% the authors and their affiliations.
%% Of note is the shared affiliation of the first two authors, and the
%% "authornote" and "authornotemark" commands
%% used to denote shared contribution to the research.
\author{David Klüner}
\email{kluener@embedded.rwth-aachen.de}
\orcid{0009-0006-3451-9435}
\authornote{Authors are shared first authors of this document.}

\author{Simon Schäfer}
\email{schaefer@embedded.rwth-aachen.de}
\orcid{0000-0002-6482-2383}
\authornotemark[1]

\affiliation{%
  \institution{RWTH Aachen, Chair for Embedded Software}
  \city{Aachen}
  \state{NRW}
  \country{Germany}
}

\author{Lucas Hegerath}
\orcid{0000-0003-2926-2664}
\email{hegerath@embedded.rwth-aachen.de}
\authornote{Authors contributed equally to this research.}

\author{Jianye Xu}
\orcid{0009-0001-0150-2147}
\email{xu@embedded.rwth-aachen.de}
\authornotemark[2]

\author{Julius Kahle}
\orcid{0000-0003-3986-1986}
\email{kahle@embedded.rwth-aachen.de}
\authornotemark[2]

\affiliation{%
  \institution{RWTH Aachen, Chair for Embedded Software}
  \city{Aachen}
  \state{NRW}
  \country{Germany}
}

\author{Hazem Ibrahim}
\orcid{0009-0007-1911-4760}
\email{hazem.ibrahim@unibw.de}
\authornotemark[2]

\affiliation{%
  \institution{University of the Bundeswehr Munich, Department of Aerospace Engineering}
  \city{Munich}
  \state{Bavaria}
  \country{Germany}
}

\author{Alexandru Kampmann}
\orcid{0009-0008-8340-1913}
\email{kampmann@embedded.rwth-aachen.de}
\affiliation{%
  \institution{RWTH Aachen, Chair for Embedded Software}
  \city{Aachen}
  \state{NRW}
  \country{Germany}
}

\author{Bassam Alrifaee}
\orcid{0000-0002-5982-021X}
\email{alrifaee@embedded.rwth-aachen.de}
\affiliation{%
  \institution{University of the Bundeswehr Munich, Department of Aerospace Engineering}
  \city{Munich}
  \state{Bavaria}
  \country{Germany}}

%%
%% By default, the full list of authors will be used in the page
%% headers. Often, this list is too long, and will overlap
%% other information printed in the page headers. This command allows
%% the author to define a more concise list
%% of authors' names for this purpose.
\renewcommand{\shortauthors}{Klüner and Schäfer et al.}

%%
%% The abstract is a short summary of the work to be presented in the
%% article.
\begin{abstract}
Ensuring the safe and efficient operation of \acp{CAV} relies heavily on the software framework used. A software framework needs to ensure real-time properties, reliable communication, and efficient resource utilization. 
Furthermore, a software framework needs to enable seamless transition between testing stages, from simulation to small-scale to full-scale experiments.
In this paper, we survey prominent software frameworks used for in-vehicle and inter-vehicle communication in \acp{CAV}. We analyze these frameworks regarding opportunities and challenges, such as their real-time properties and transitioning capabilities. Additionally, we delve into the tooling requirements necessary for addressing the associated challenges. We illustrate the practical implications of these challenges through case studies focusing on critical areas such as perception, motion planning, and control. Furthermore, we identify research gaps in the field, highlighting areas where further investigation is needed to advance the development and deployment of safe and efficient \ac{CAV} systems.
\end{abstract}

%%
%% The code below is generated by the tool at http://dl.acm.org/ccs.cfm.
%% Please copy and paste the code instead of the example below.
%%

\begin{CCSXML}
<ccs2012>
   <concept>
  <concept_id>10010520.10010553.10010554.10010556</concept_id>
  <concept_desc>Computer systems organization~Robotic control</concept_desc>
  <concept_significance>300</concept_significance>
  </concept>
   <concept>
  <concept_id>10011007.10010940.10010971.10010972</concept_id>
  <concept_desc>Software and its engineering~Software architectures</concept_desc>
  <concept_significance>500</concept_significance>
  </concept>
   <concept>
  <concept_id>10010147.10010178.10010219.10010220</concept_id>
  <concept_desc>Computing methodologies~Multi-agent systems</concept_desc>
  <concept_significance>500</concept_significance>
  </concept>
 </ccs2012>
\end{CCSXML}

\ccsdesc[300]{Computer systems organization~Robotic control}
\ccsdesc[500]{Software and its engineering~Software architectures}
\ccsdesc[500]{Computing methodologies~Multi-agent systems}

\keywords{Software Frameworks, Software Architectures, Small-Scale Testbeds, Full-Scale Testbeds, Simulations, Automated Vehicles, Middlewares}

\received{31 March 2025}
% \received[revised]{12 March 2009}
% \received[accepted]{5 June 2009}

%%
%% This command processes the author and affiliation and title
%% information and builds the first part of the formatted document.
\maketitle

\section{Introduction}
\label{sec:intro}

\acp{CAV} promise new control, planning and cooperation paradigms for vehicles in the future \cite{guanetti_control_2018}. 
They represent one of the newest trends to achieve automated operation without driver involvement \cite{burkacky_rethinking_nodate}.
In contrast to regular vehicles, \acp{CAV} posses the capability of communication with other vehicles and infrastructure and offers the option to cooperate during operation.
This cooperation could enable the softening of driving laws, as \acp{CAV} could plan their paths jointly, removing the necessity for rigid rules when operating a vehicle \cite{guanetti_control_2018, elliott_recent_2019}.
\acp{CAV} equally enable seamless updates of software for deployed vehicles, offering new commercial opportunities.
These advancements are enabled by new communication technologies and improvements in software. 
Automated operation, inter-vehicle communication, and the increasing demands on vehicle functions lead to more compute capacity and generally more software in vehicles.
Software is rarely built in isolation, but is often based on existing frameworks \cite{burkacky_rethinking_nodate, wang_review_2024}. 
Software frameworks provide domain-specific solutions, for example execution of \ac{ML} models, and consequently, we believe that their importance in next-generation vehicles will increase.
As a result, the safe and efficient operation of \acp{CAV} is critically dependent on the software architecture and frameworks used \cite{autosar_explanationPlatformDesign_2024}. 

The \textit{software architecture} is defined as the gross structure of a software system \cite{garlan_software_2000}. 
Of particular interest in the software architecture of an \system\ is the middleware \cite{kluner_modern_2024}.
It has emerged as a core component of \acp{CAV} and future vehicles and serves as a foundational framework \cite{zhu_requirements-driven_2021}.
Automotive middlewares function as intermediate communication layers between the application and the system software.
They provide the core functions of communication and structured software execution in distributed, heterogeneous automotive \ac{E/E} architectures \cite{neely_adaptive_2006, kluner_modern_2024}. 

\begin{figure}[tbp]
    \centering
    \includegraphics[width=\linewidth]{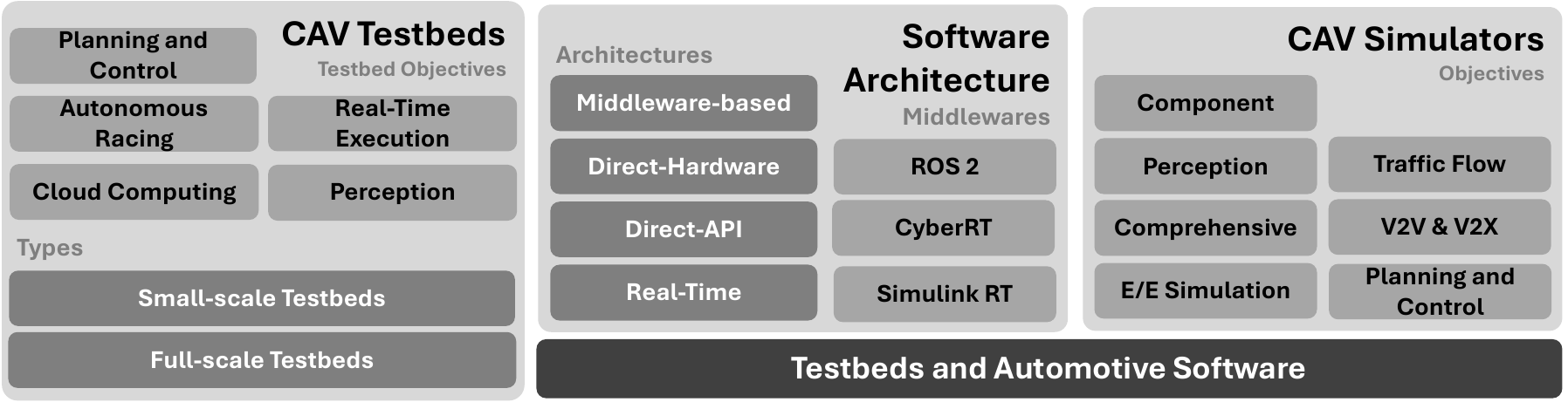}
    \caption{The domains of this review: Testbeds, simulators and software frameworks and their overlapping domains.}
    \Description{The domains of this review: Testbeds, simulators and software frameworks and their overlapping domains.}
    \label{fig:overview}
\end{figure}

Validating the correct operation of software testing is required.
In the \ac{CAV} domain, many approaches are initially tested on smaller scales or within simulation environments before being deployed on a full scale. 
Smaller-scale testing is often preferred due to its cost-effectiveness, the feasibility of easily testing with multiple vehicles, and the minimal consequences of failures \cite{mokhtarian2024survey}. 
Simulations provide an even cheaper alternative, but at the cost of reduced accuracy. 
The effectiveness of a simulation is limited by the quality of its implementation and the specific objectives for which it is designed \cite{li_choose_2024}. 
The goal is often to achieve the full-scale applicability of novel methods.
In this review, we examine the intersection of software architecture and CAV testbeds, what architectures are suitable for which test case, and what should be considered in the implementation of novel concepts when planning for an evaluation. 
\cref{fig:overview} illustrates the concepts we investigated in this survey and the topics in each domain.

\subsection{Main Contributions}
\label{sec:intro:contributions}

The main contributions of this survey include:
\begin{itemize}
    \item We introduce the domain of automotive software frameworks, software architecture and test cases for \acp{CAV}.
    \item We present a survey of simulators, small-scale testbeds, and full-scale testbeds.
    \item We derived requirements for transitioning software from simulation to small-scale to full-scale testing.
    \item We provide recommendations on best practices for testing software for \acp{CAV}.
\end{itemize}

\subsection{Outline}
\label{sec:intro:outline}
 \cref{sec:background} provides an overview of the foundational background on \acp{CAV}, software architectures, automotive software stacks and middleware. \cref{sec:simulation} presents a survey of automotive simulation approaches and their associated software architectures. 
 \cref{sec:small-scale} examines small-scale testbeds, including their architectures and middlewares.
The requirements imposed by small-scale testbeds on the evaluated approaches are detailed in \cref{sec:Requirements:small-scale}. 
\cref{sec:full-scale} focuses on full-scale testbeds, their architectures, and employed middlewares. 
The specific requirements for testing at full scale are addressed in \cref{sec:Requirements:full-scale}. 
Finally, \cref{sec:transitioning} explores software that bridges small- and full-scale testing.

\section{Definitions}
\label{sec:background}

This section introduces background information necessary to understand the main focus of this paper. 
First, we define the differences between \acp{CAV} and regular vehicles, then we discuss automotive software stacks and the role of the middleware as a data distributor.
Lastly, we discuss testing methods for \ac{CAV} ranging from \simulators\ to full-scale experiments with real vehicles.

% \begin{figure}[bt]
%\centering
%\includegraphics[width=1.0\linewidth]{figures/Intro.pdf}
%\caption{Testbeds, use cases and images from simulators, small-scale testbeds and full-scale testbeds. Simulators depict Sumo and Carla, as a small-scale example the CPM Lab. Images from: \cite{noauthor_sumo_nodate, dosovitskiy_carla_2017, dosovitskiy_carla_2017, kessler_bridging_2019}.}
%\Description{}
%\label{fig:intro}
% \end{figure}

\subsection{Connected and Automated Vehicle}
\label{sec:background:cav}

Current research focuses on interconnecting vehicles to achieve objectives such as platooning of vehicles and automating their driving functions to reduce driver load and enable novel use cases.
We follow the definition of \citeauthor{guanetti_control_2018} in \cite{guanetti_control_2018} and refer to these vehicles as \acfp{CAV}. 
According to their definition, \acp{CAV} are capable of automated driving and connectivity with other road users. 
Automated driving involves automated operation, ranging from operation under perfect conditions to full autonomy without user intervention \cite{guanetti_control_2018}.
Communication is realized commonly using \ac{V2X} protocols, which enables vehicles to exchange limited information with other entities \cite{kueppers_v2aix_2024}. A common example of this communication is the \ac{V2V} communication, where vehicles share information about themselves with other vehicles in their immediate vicinity. 

\subsection{Software Architecture}
\label{sec:background:softarch}

The software architecture, as defined by \citeauthor{garlan_software_2000} in \cite{garlan_software_2000}, is the overall structure of a software system.
It outlines the components, their essential properties, and the interactions between them.
These components can vary in form, such as classes within an application, services for a web application, or even entire applications that constitute a distributed system.
Additionally, the architecture specifies the ways in which these components interact with each other.
For instance, classes can interact through function calls, while web applications may utilize HTTPS API calls.
By defining these interactions, the software architecture ensures a clear division of responsibilities and pathways within the system. 

\subsection{Software Framework}
\label{sec:background:frameworks}

Software frameworks provide mechanisms for common or recurring challenges, such as message exchange or deep learning applications. 
Frameworks commonly have reusable implementations or entire components that multiple applications may reuse \cite{bass_software_2012}.
In the automotive domain, we distinguish between automotive middlewares and domain-specific frameworks.
Automotive middlewares focus on software architecture and communication between components in the vehicle, while other frameworks focus on specific domains, for example, environmental perception, control, or machine learning applications \cite{kluner_modern_2024}. 

\subsection{Automotive Middlewares}
\label{sec:background:middlewares}

\begin{figure}[tbp]
    \centering
    \includegraphics[width=\linewidth]{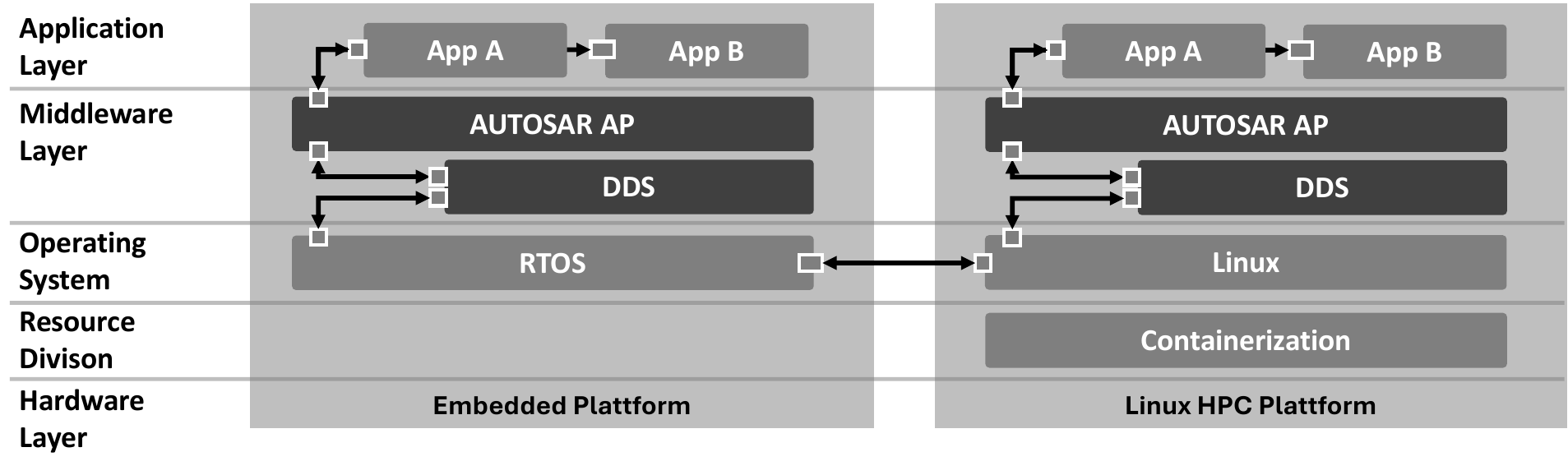}
    \caption{The role of a middleware in the development of distributed software. The middleware functions as an intermediary between the application and the operating system. In this case AUTOSAR Adaptive makes use of \ac{DDS} and the operating systems network stack to communicate across the two devices.}
    \Description{The role of a middleware in the development of distributed software. The middleware functions as an intermediary between the application and the operating system. In this case AUTOSAR Adaptive makes use of \ac{DDS} and the operating systems network stack to communicate across the two devices.}
    \label{fig:middleware_idea}
\end{figure}

Automotive Middlewares form the core of an automotive software architecture. 
Implemented as a software framework, they act as intermediate communication layers between the operating system and the application layer \cite{neely_adaptive_2006}.
As a result, the middleware decouples the software from the precise underlying hardware and network topology.
Using new communication patterns on automotive networks, they interconnect vehicles in new \ac{E/E} architecture.
Middlewares differ in scope. 
While some middlewares, also called architecture platforms, accomplish comprehensive functions, others contribute primarily to communication.
Architecture platforms address multiple domains in automated vehicles, such as communication, security, resource control, scheduling and execution \cite{henle_architecture_2022}, while others such as FastDDS address mainly communication \cite{eprosima_11_2024}.
\Cref{fig:middleware_idea} illustrates an example use case of the automotive middleware AUTOSAR Adaptive and how it connects two compute units running different operating systems and applications.

\subsection{Test Automotive Software} 
\label{sec:background:testbeds}
Testing software is a critical aspect of the development process, ensuring that systems function as intended and are robust against various potential failures. 
Several methodologies exist for testing software, ranging from automatic test generation to unit tests, integration tests, and other classifications such as black-box, white-box, and gray-box testing \cite{tang_survey_2023}.
These approaches differ in their scope, the level of detail they provide, and the specific aspects of the software or system they target.

For researchers in the automotive domain, two key factors, realism and cost, often influence the choice of testing approach \cite{tang_survey_2023}.
A central question for research groups is how realistic the test setup can be, especially when considering the need for high-detail results that reflect the behavior of automotive systems.
 Increased realism often increases the cost of a test setup. 

One side of this trade-off is performing tests on full-scale vehicles. 
Although this offers the highest degree of realism, it is also the most expensive option, both in terms of the resources required to maintain real vehicles and the potential risks involved if failures occur during testing. 
Full-scale testing also requires significant space and infrastructure, making it impractical for early-stage or frequent testing.

Given these requirements, three primary approaches to automotive software testing are commonly used, each offering a different balance of realism, cost, and feasibility:

\begin{enumerate}
    \item Simulated environments allow for highly flexible testing scenarios, with adjustable levels of fidelity depending on the specific aims of the tests. This approach can range from basic \simulators\ of vehicle dynamics to highly detailed models that closely mimic real-world behavior. The advantage of simulation lies in its scalability and the ability to explore a wide range of conditions, but it may not capture certain real-world details \cite{kaur_survey_2021}. In particular, simulation of vehicle dynamics and sensor behavior is an ongoing challenge.
    \item Testing in Small-scale testbeds, this approach uses miniature scaled vehicles, which offer a lower-cost and lower-complexity alternative to full-scale testing. Although the scale may introduce certain approximations in terms of dynamics, it provides a practical method for early-stage testing and validation without the need for large test environments or significant financial investment \cite{mokhtarian2024survey}.
    \item Testing on actual vehicles represents the most accurate approach, capturing real-world variables and complexities involved. However, this method is often prohibitive in terms of cost, space requirements, and associated risks. For research groups, this method is often not feasible and more commonly employed commercially. Although it provides the highest degree of fidelity, its use in research is typically limited to later stages of development or when smaller-scale or simulated approaches are insufficient \cite{yurtsever_survey_2020}.
\end{enumerate}

\section{Simulation}
\label{sec:simulation}

Simulations are central to researching and testing \acp{CAV}.
They offer a controlled and often simplified environment in which researchers and developers can conduct repeatable experiments, providing an accessible approach for testing and validation.
This section introduces the concept of simulation for \acp{CAV}, discusses simulator categories, and examines the middleware used across various simulators.
We first outline commonly found categories of simulators and extend these with additional categorizations based on our own findings.
Next, we present and analyze the middleware technologies that underlie these simulators.

\subsection{Introduction}
\label{sec:simulation:introduction}

Much like small-scale testbeds, simulators frequently simplify their environment to focus on particular subdomains within CAV research.
For example, traffic simulators such as SUMO primarily model vehicle behavior at a macroscopic level but do not provide high-fidelity simulations of the vehicle environment \cite{noauthor_sumo_nodate}.
In contrast, simulators like CARLA offer a detailed vehicle-perception view, capturing nuances of vehicle sensing and surrounding conditions \cite{dosovitskiy_carla_2017}.

In such simulation environments, developers create models of vehicle behavior that reflect specific aspects of real-world driving.
Depending on the use case, only limited fidelity may be required or desired \cite{li_choose_2024}, as creating and maintaining complex simulation models can prove expensive.
For instance, a bicycle model can suffice to represent vehicle dynamics for certain experimental purposes \cite{polack_kinematic_2017}.

Simulations can be performed at a relatively low cost and effort, are readily accessible through open-source platforms, and scale effectively to accommodate large numbers of vehicles.
They provide a controlled environment where repeatable experiments can be conducted, ensuring that the results are reproducible and shareable \cite{tang_survey_2023}.
However, the development of high-fidelity simulators entails significant effort.
Furthermore, fidelity constraints mean certain real-world complexities cannot be fully replicated, and scaling up to detailed simulations may require substantial computational resources.

Finally, as part of our literature review, we identified several simulators that are listed in \cref{tab:simulators-overview}.
They are loosely grouped based on their primary focus, though many simulators also offer functionality beyond their key domain.
The following subsection categorizes these simulators in more detail and explores the middleware technologies that underpin their operation.

\subsection{Categories of Simulators}
\label{sec:simulation:simulators}

\begin{figure}[tbp]
    \centering
    \includegraphics[width=\linewidth]{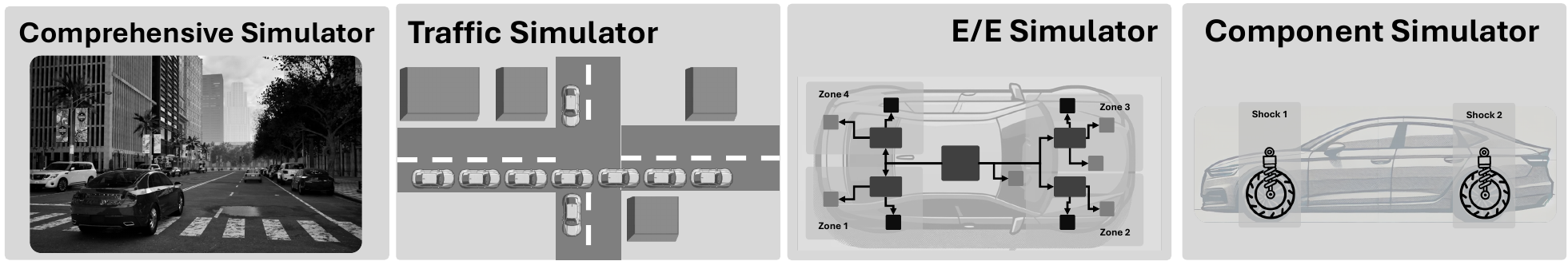}
    \caption{Illustration of the classifications for simulations. We found four classifications: comprehensive simulator, traffic flow simulator, E/E simulator, and component simulator. Image of comprehensive simulator from \cite{team_carla_nodate}.}
    \Description{Illustration of the classifications for simulations. We found four classifications: comprehensive simulator, traffic flow simulator, E/E simulator, and component simulator. Image of comprehensive simulator from [219].}
    \label{fig:simulators}
\end{figure}

% Simulator - Source - Class - used by 
\begin{table*}[tbp] 
\caption{Overview of the simulators we found during our literature review. The simulators are loosely grouped based on their primary focus, and they may also provide functionality beyond their primary focus.}
\Description{Overview of the simulators we found during our literature review. The simulators are loosely grouped based on their primary focus, and they may also provide functionality beyond their primary focus.}
\label{tab:simulators-overview}
\centering
\resizebox{\linewidth}{!}{
\begin{tabular}{llllll}
\hline
\textbf{Simulator} & \textbf{Year} & \textbf{OSS} & \textbf{Middleware} & \textbf{Focus}   & \textbf{Related Publication} \\
\hline

\multicolumn{6}{l}{\textbf{Traffic Simulator}}\\
SUMO\cite{noauthor_sumo_nodate} & 2001 & \checkmark & None & Traffic Simulator & \cite{nguyen_towards_2020, alvarez_lopez_microscopic_2018} \\
TransModeler\cite{noauthor_transmodeler_nodate} & 2005 & \texttimes & None & Traffic Simulator & \cite{li_longitudinal_2022}\\
AutonoVISim\cite{best_autonovi-sim_2018} & 2018 & \texttimes & None & Traffic Simulator & -\\
CityFlow \cite{zhang_cityflow_2019} & 2019 & \checkmark & None & Traffic Simulator & - \\
LGSVL\cite{rong_lgsvl_2020} & 2020 & \texttimes & None & Traffic Simulator & \cite{antonante_monitoring_2021} \\
POLARIS\cite{noauthor_polaris_2021} & 2021 & \texttimes & None & Traffic Simulator & \cite{juszkiewicz_use_2023} \\
V2XSim \cite{li_v2x-sim_2022} & 2022 & \texttimes & None &V2X Communication & -\\
CAVsim \cite{zhang_cavsim_2023} & 2023 & \texttimes & None & Traffic Simulator & \cite{he_comparative_2023}\\
LimSim \cite{wen_limsim_2023} & 2023 & \texttimes & None & Traffic Simulator & \cite{wen_trafficmcts_2023} \\
Nocturne \cite{vinitsky_nocturne_2023} & 2023 & \texttimes & None & Traffic Simulator & \cite{muthali_multi-agent_2023} \\

\midrule 
\multicolumn{6}{l}{\textbf{Component Simulator}}\\
MATLAB Simulink\cite{matlab2024simulink} & 1984 & \texttimes & None & Component & \cite{zhang_design_2022, soma_simulation_2016, basciani_adas_2023} \\
ApolloSim \cite{kepets_apollosim_2024} & 2024 & \checkmark & None & Component & \cite{ai_ws-3d-lane_2023} \\

Anylogic \cite{noauthor_anylogic_nodate} & 2000 & \texttimes & ROS & Component & \cite{bock_analytical_2017}\\

\midrule 
\multicolumn{6}{l}{\textbf{E/E Simulator}}\\
CarMaker \cite{noauthor_carmaker_nodate} & 1999 & \texttimes & Proprietary & E/E Simulator & \cite{li_real-time_2021} \\
FERAL\cite{kuhr_feral_2013} & 2013 & \texttimes & Proprietary & E/E Simulator & \cite{bachorek_towards_2019} \\
MOSAIC \cite{fokus_eclipse_nodate} & 2023 & \texttimes & Proprietary & E/E Simulator & \cite{schrab_modeling_2023}\\

\midrule 
\multicolumn{6}{l}{\textbf{Comprehensive Simulator}}\\
VISSIM \cite{noauthor_traffic_nodate} & 1992 & \checkmark & None & Comprehensive Simulator & \cite{lownes_vissim_2006} \\
TBSim\cite{xu_bits_2022} & 2006 & \texttimes & None & Comprehensive Simulator & -\\
Nvidia Drive Sim\cite{noauthor_nvidia_nodate} & 2016 & \texttimes & None & Comprehensive Simulator & \cite{popov_mitigating_2024} \\ 
DeepDrive \cite{noauthor_deepdrive_nodate} & 2018 & \checkmark & None & Comprehensive Simulator & -\\
CiThruS2 \cite{galazka_cithrus2_2021} & 2021 & \checkmark & None & Comprehensive Simulator & -\\
InterSim \cite{sun_intersim_2022} & 2022 & \checkmark & None & Comprehensive Simulator & -\\
MetaDrive \cite{li_metadrive_2022} & 2022 & \checkmark & None & Comprehensive Simulator & \cite{mo_freecontrol_2024} \\
L2R \cite{herman_learn--race_2021} & 2022 & \checkmark & None & Comprehensive Simulator & \cite{francis_learn--race_2022} \\
Waymax \cite{gulino_waymax_2023} & 2023 & \checkmark & None & Comprehensive Simulator & \cite{xiao_easychauffeur_2024} \\

Gazebo \cite{noauthor_gazebo_nodate} & 2012 & \checkmark & ROS 2 & Comprehensive Simulator & \cite{syed_ahamed_software---loop_2018} \\
BEAMNG \cite{noauthor_home_2024} & 2015 & \texttimes & ROS 2 & Comprehensive Simulator & \cite{basciani_adas_2023} \\
CARLOS\cite{geller_carlos_2024} & 2024 & \checkmark & Proprietary / ROS 2 & Comprehensive Simulator & - \\
CARLA\cite{dosovitskiy_carla_2017} & 2017 & \checkmark & Proprietary & Comprehensive Simulator & \cite{brogle_hardware---loop_2019} \\
AutoDrive \cite{samak_autodrive_2021} & 2022 & \checkmark & Proprietary & Comprehensive Simulator & \cite{samak_autodrive_2021}\\
\hline
\end{tabular}
}
\end{table*}

A core advantage of using simulators lies in the simplified environment they create, allowing researchers and developers to focus on specific research questions. For instance, in motion-planning studies, a simulator may centrally track vehicle positions so that the principal challenge, planning the vehicle’s motion, can be emphasized. We discovered a diverse ecosystem of CAV simulators, which \citeauthor{li_choose_2024} \cite{li_choose_2024} divide into five classes: traffic flow simulators, sensory data simulators, driving policy simulators, vehicle dynamics simulators, and comprehensive simulators. In addition, our findings revealed two more categories not mentioned by \citeauthor{li_choose_2024} \cite{li_choose_2024}: E/E simulators and component simulators. Altogether, we consider the following seven classes of CAV simulators:

\begin{itemize} 
    \item \textbf{Traffic Flow Simulators.}
    These simulators operate on the largest scale and model microscopic vehicles with independent behavior, although each vehicle’s behavior can be simplified (e.g., SUMO \cite{noauthor_sumo_nodate}, TransModeler \cite{noauthor_transmodeler_nodate}). Their primary applications include \acp{CAV} development, road infrastructure planning, and vehicle-to-everything technology development \cite{li_v2x-sim_2022}. They typically run on a single machine and are tightly integrated with the vehicle behavioral software for performance reasons.
    
    \item \textbf{Sensory Data Simulators.}  
    Sensory data simulators aim to generate realistic sensor data and ground-truth information for a single vehicle (e.g., CARLA \cite{dosovitskiy_carla_2017}). They emphasize high fidelity in both visual and sensor outputs, while other vehicles in the simulation are often modeled with limited accuracy (e.g., following preplanned paths). These simulators are particularly useful for developing full-stack autonomous driving and perception systems, as they provide detailed sensor data.
    
    \item \textbf{Driving Policy Simulators.}  
    Driving policy simulators involve large-scale environments containing multiple vehicles, each operating as an independent agent following its own driving policy. They are especially valuable for developing and evaluating motion-planning algorithms (e.g., CAVSIM \cite{zhang_cavsim_2023}).
    
    \item \textbf{Vehicle Dynamics Simulators.}  
    These simulators focus on accurately recreating vehicle dynamics, often simplifying behavior to follow physical laws. Their primary application is in vehicle dynamics and control system development (e.g., CarMaker \cite{noauthor_carmaker_nodate}).
    
    \item \textbf{Comprehensive Simulators.}  
    Comprehensive simulators combine multiple aspects—planning, control, and perception—into a unified platform. Examples include CARLA \cite{dosovitskiy_carla_2017}, BeamNG \cite{noauthor_home_2024}, Gazebo \cite{noauthor_gazebo_nodate}, and CiThruS2 \cite{galazka_cithrus2_2021}. These often employ middlewares such as \ac{ROS 2}% or Apollo CyberRT \cite{macenski_robot_2022, noauthor_apollo_nodate}
    , making them well-suited for testing software stacks with minimal adjustments. Though broad in scope, some comprehensive simulators have limited support for large-scale traffic simulation.
    
    \item \textbf{E/E Simulators.}  
    A separate category of simulators focuses on electric/electronic (E/E) architectures, such as CarMaker \cite{noauthor_carmaker_nodate}, MOSAIC \cite{fokus_eclipse_nodate}, and FERAL \cite{kuhr_feral_2013}. Mostly industry-developed and proprietary, these tools simulate aspects of the vehicle’s internal E/E systems, which can complicate integration into open research frameworks due to restricted visibility into their internals.
    
    \item \textbf{Component Simulators.}  
    Finally, there exist simulators that target a single hardware component rather than the entire vehicle, such as MATLAB Simulink \cite{zhang_design_2022, soma_simulation_2016} and AnyLogic \cite{noauthor_anylogic_nodate}. These focus on lower-level physical phenomena, typically operating independently of a higher-level software stack. As a result, they occupy a more specialized niche and remain orthogonal to full-stack CAV framework evaluations.
\end{itemize}

We identified several simulators within each class that support \ac{CAV} systems. \Cref{tab:simulators-overview} provides an overview of our findings, including the two additional categories (E/E and component simulators) not described by \citeauthor{li_choose_2024} \cite{li_choose_2024}. \Cref{fig:simulators} illustrates these simulator classes and offers an intuitive sense of how they align with various CAV research needs. Notably, we observed many industry-developed, closed-source simulators whose proprietary nature poses challenges for the independent evaluation of software frameworks.

\subsection{Architectures and Middlewares of Simulators}
\label{sec:simulation:architectures}

In addition to categorizing simulators according to their primary focus (see \cref{sec:simulation:simulators}), we also analyzed how the simulators connected to the software under test. 
While we did not classify the simulator’s internal architecture, we examined whether the software under test interacted directly with the simulator’s \ac{API} or through a middleware framework.
\Cref{tab:simulators-overview} highlights the architectures implemented by each simulator discussed.

Some simulators, especially those centered on traffic flow or component, level modeling—use a direct connection between software components.
In these cases, no middleware is employed, as it is generally unnecessary for component simulators and can introduce inefficiencies in traffic-flow simulations.
Additionally, many vehicle simulation models represent only basic vehicle behavior, eliminating the need for complex communication middleware.

\begin{itemize}
    \item An example for an \ac{API}-based simulator is CARLA, a prominent Full-Stack simulator. The simulator offers a Python interface for direct control of the simulation; as a result, software interfaces directly with the simulator instead of relying on middleware \cite{dosovitskiy_carla_2017}. This simulator was extended in \cite{geller_carlos_2024} to create a new CARLOS framework. CARLOS exposes CARLA’s proprietary API to the ROS 2 ecosystem. While this extension is a bridge, it is not a proper middleware since the simulation’s internals remain unchanged. Instead, it transitions communication between the two systems through mapping API calls to \ac{ROS 2} topics and service calls.
    \item MATLAB Simulink is a proprietary programming environment enabling numerical computation, data analysis, and algorithm development \cite{matlab2024simulink}. Users can model and simulate physical systems through its Simulink toolbox, which includes individual vehicle components. Although Simulink is not a middleware solution, it can interface with other platforms (e.g., \ac{ROS 2}) for communication and control purposes. Users benefit from a graphical interface, extensive documentation, and a large community. MATLAB, however, requires a commercial license.
\end{itemize}

By contrast, comprehensive simulators that target full vehicle software stacks in distributed configurations often rely on middleware-based architectures. These architectures more closely resemble actual deployment environments. Examples of common middleware or software frameworks in these setups include \acf{ROS} and \acf{ROS 2}.

\begin{itemize}
    \item \ac{ROS 2} is a software framework for real-time robotic application development \cite{macenski_2022_ros2}.
    It uses a node-based computational graph, wherein each node represents a discrete software component.
    Topics enable anonymous communication by letting senders and receivers share only a common topic name \cite{open_robotics_quality_2024}.
    Task execution in \ac{ROS 2} is managed through the executor model, which orders callbacks, timers, and other asynchronous events \cite{open_robotics_executors_2024}. 
    While deployment capabilities largely involve in-place compilation, \ac{ROS 2} has a large open-source community and a robust ecosystem of industrial support (e.g., NVIDIA, Intel).
    \item Often referred to as \ac{ROS}, the original Robot Operating System provides a similarly node-based computational graph but relies on a master node for service registration and topic coordination.  
    It remains widely used in research and industry, supported by an extensive open-source community and a vast repository of packages, even though the framework reached end-of-life.  
    However, \ac{ROS} does not offer native real-time implementations or multi-distribution support.  
    Its networking model also differs from \ac{ROS 2}, making large-scale distributed deployments more challenging.  
    Despite these limitations, many simulation tools and frameworks, from older versions of Gazebo to specialized navigation stacks, remain built around \ac{ROS}.
\end{itemize}

\section{Small-Scale Testbeds}
\label{sec:small-scale}

Small-scale testbeds serve as an intermediary between simulation and full-scale testbeds. They provide a controlled and often simplified environment for researchers and developers to conduct experiments with a wider range of realms compared to simulations.
Similar to Section \ref{sec:simulation}, this section introduces the concept of small-scale testbeds, explores various categories, and examines the middleware used across different testbeds.
We begin by defining a small-scale testbed and presenting a collection of different testbeds. Next, we analyze the middleware technologies employed by these testbeds and introduce technologies not previously discussed in Section \ref{sec:simulation:architectures}.

\begin{figure}[tbp]
    \centering
    \includegraphics[width=\linewidth]{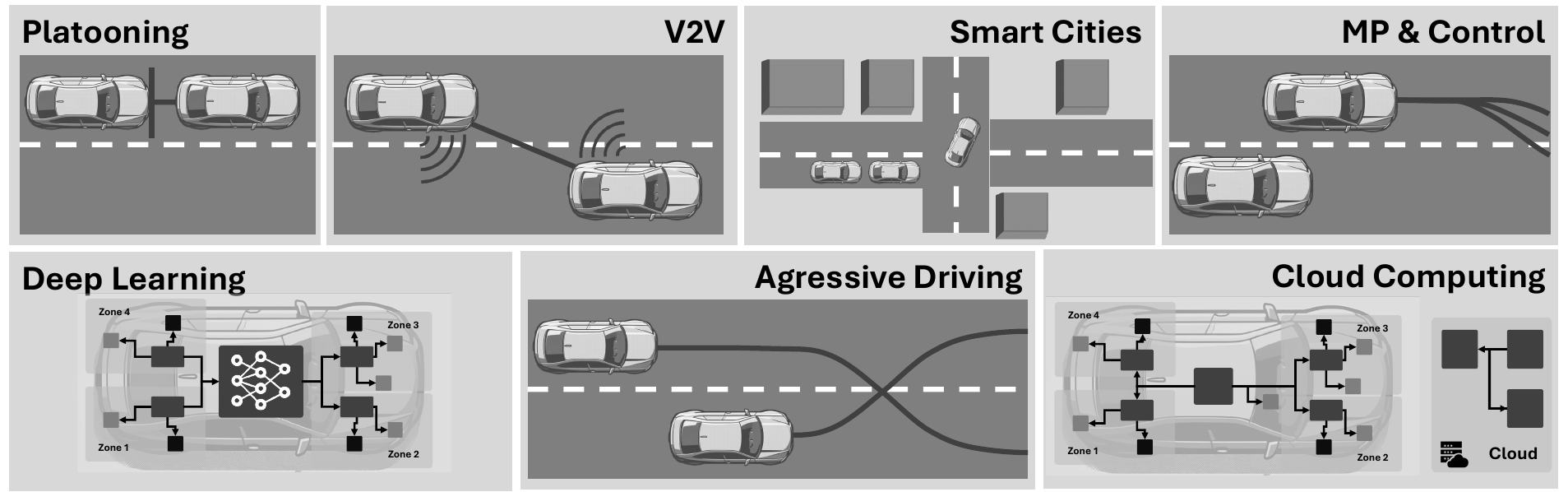}
    \caption{Illustration of the classifications for small-scale testbeds. We found six classifications: platooning, V2V communication, smart cities, motion planning \& control, deep learning, aggressive driving, and cloud computation.}
    \label{fig:small-scale-testbeds}
    \Description{Illustration of the classifications for small-scale testbeds. We found six classifications: platooning, V2V communication, smart cities, motion planning \& control, deep learning, aggressive driving, and cloud computation.}
\end{figure}

Small-scale testbeds act as an intermediary between simulation and full-scale testbeds, providing a controlled physical environment to validate concepts and technologies in \acp{CAV} \cite{scheffe_scaled_2023, schafer2023investigating}. 
These testbeds typically use scaled-down vehicles or platforms that simulate real-world dynamics and interactions, mitigating the cost and safety concerns associated with full-scale experiments. 
They bridge the gap between virtual and real-world experiments by offering physical presence and interactivity, which are absent in simulations.
The software architecture for small-scale experiments sometimes mirrors those used in full-scale implementations but on a smaller scale. This includes the use of middleware like \ac{ROS 2} for coordinating distributed subsystems, as well as real-time computation frameworks to evaluate algorithms under timing constraints. 

\begin{table*}[tbp]
\centering
\caption{Overview of the small-scale testbeds we found during our literature review. The testbeds are loosely grouped based on their primary focus, and they may also provide functionality beyond their primary focus.}
\label{tab:small-scale-testbeds}
\resizebox{\linewidth}{!}{%
\begin{tabular}{llllllll}
\toprule
\textbf{Testbed} & \textbf{Year} & \textbf{Scale} & \textbf{OSS} & \textbf{Middleware} & \textbf{Focus} & \textbf{Related Publication} \\
\midrule

\multicolumn{7}{l}{\textbf{Platooning}}\\
RoboCoPlat \cite{filho2020cooperative}  & 2020  & 1:10  & \texttimes  & \ac{ROS}  & Platooning / Communication  & \cite{filho2021copadrive, filho2023evaluation}  \\
Cyclops \cite{lee2022cyclops}  & 2022  & 1:14  & \checkmark  & \ac{ROS}  & Platooning  & \cite{koo2023phalanx}  \\
\citeauthor{dejager2022autonomous} \cite{dejager2022autonomous}  & 2022  & -  & \texttimes  & \ac{ROS}  & Platooning  & -  \\ 
\citeauthor{yu2021design} \cite{yu2021design}  & 2021  & 1:20  & \texttimes  & \ac{ROS}, MicroROS  & Autonomous Parking  & -  \\

\midrule
\multicolumn{7}{l}{\textbf{V2V}}\\
SVEA \cite{jiang2022svea}  & 2022  & 1:10  & \checkmark  & \ac{ROS}  & Communication  & \cite{chen2023safe, yu2024online}  \\
\citeauthor{elmoghazy2024realworld} \cite{elmoghazy2024realworld}  & 2024  & 1:10  & \texttimes  & \ac{ROS}  & Communication  & -  \\
\citeauthor{krieger2023integration} \cite{krieger2023integration}  & 2023  & 1:10  & \texttimes  & \ac{ROS 2}  & 6G Communication  & \cite{priyanta2024it}  \\
Go-CHART \cite{kannapiran2020gochart}  & 2020  & 1:28  & \checkmark  & \ac{ROS}  & Perception  & -  \\
\citeauthor{ehrenfeuchter2024realtime} \cite{ehrenfeuchter2024realtime}  & 2024  & 1:10  & \texttimes  & \ac{ROS}  & Perception  & -  \\
\citeauthor{jaimes2019lowcost} \cite{jaimes2019lowcost}  & 2019  & 1:18  & \texttimes  & \ac{ROS}  & Localization  & \cite{ibarra2021application}  \\
CARTRAC \cite{babu2022comprehensive}  & 2022  & (1:15)  & \texttimes  & - (Wi-Fi)  & Communication  & \cite{babu2022dtmr}  \\
STARS \cite{lodato2018scaled}  & 2018  & (1:9)  & \texttimes  & - (Wi-Fi)  & Sensor Technologies  & -  \\
\citeauthor{weinbauer2023taking} \cite{weinbauer2023taking}  & 2023  & -  & \checkmark  & - (Wi-Fi)  & Communication  & -  \\
RMS \cite{chen2020handson}  & 2019  & 1:19  & \texttimes  & \ac{ROS}  & Education (Middle School) / Computer Vision  & -  \\

\midrule
\multicolumn{7}{l}{\textbf{Smart Cities}}\\
IDS3C (UDSSC) \cite{stager2018scaled}  & 2018  & 1:25  & \texttimes  & \ac{ROS}  & Smart City / Planning and Control  & \cite{chalaki2022research, zayas2022digital} \\
Duckietown \cite{paull2017duckietown}  & 2017  & (1:13)  & \checkmark  & \ac{ROS}  & Smart City / Planning and Control  & \cite{cheng2018integration, tani2017duckietown}  \\
MiniCity \cite{buckman2022evaluating}  & 2022  & 1:10  & \checkmark  & \ac{ROS}  & Smart City / Perception  & \cite{buckman2023infrastructurebased, murali2024learning}  \\
Miniature Autonomy \cite{tiedemann2022miniature}  & 2022  & 1:87  & \checkmark  & \ac{ROS}  & Smart City / Planning and Control  & \cite{tiedemann2024miniature, pareigis2023improved}  \\
ICAT \cite{tian2024icat}  & 2024  & (1:15)  & \checkmark  & \ac{ROS}  & Smart City / Planning and Control  & \cite{he2024advanced}  \\
MRS Smart City \cite{mahdi2024mrs}  & 2024  & -  & \texttimes  & \ac{ROS}  & Smart City  & -  \\
DonkieTown \cite{larralde-ortiz2023donkietown}  & 2023  & 1:10  & \checkmark  & \ac{ROS}  & Smart City / Planning and Control  & -  \\
\citeauthor{zdesar2023cyberphysical} \cite{zdesar2023cyberphysical}  & 2023  & -  & \texttimes  & \ac{ROS}  & Smart City / Localization / Perception  & -  \\
\citeauthor{vargas2024design} \cite{vargas2024design}  & 2024  & 1:10  & \texttimes  & \ac{ROS}  & Smart City / Mixed Traffic / Computer Vision  & -  \\
\citeauthor{morrissett2019physical} \cite{morrissett2019physical}  & 2019  & 1:18  & \texttimes  & - (Radio)  & Smart City / Traffic Management  & -  \\
CyPhyHouse \cite{ghosh2020cyphyhouse}  & 2020  & 1:10  & \checkmark  & Custom (\ac{ROS}-Based)  & Aerial and Ground Vehicles  & \cite{ghosh2018language, hsieh2021skytrakx}  \\
TAM-T \cite{stauffer2023tactical}  & 2023  & 1:10  & \checkmark  & - (Radio)  & Aerial and Ground Vehicles  & -  \\

\midrule
\multicolumn{7}{l}{\textbf{MP \& Control}}\\
CPM Lab \cite{kloock_cyber-physical_2021}  & 2021  & 1:18  & \checkmark  & \ac{ROS 2} / \ac{DDS}  & Planning and Control / Mixed Traffic  & \cite{mokhtarian_cpm_2023, kloock_architecture_2023, kloock_testing_2023, scheffe_networked_2020, kloock_vision-based_2020, scheffe_scaled_2023, xu2024sigmarl, xu2024xpmarl}  \\
Cambridge RoboMaster \cite{blumenkamp2024cambridge}  & 2024  & (1:14)  & \checkmark  & \ac{ROS 2}  & Planning and Control  & -  \\
CHARTOPOLIS \cite{ulhas2022chartopolis}  & 2022  & (1:15)  & \texttimes  & \ac{ROS} / DonkeyCar  & Planning and Control / Perception  & \cite{ulhas2024ganbased}  \\
CRS \cite{carron2023chronos}  & 2023  & 1:28  & \checkmark  & \ac{ROS}  & Planning and Control / System Identification  & \cite{bodmer2024optimizationbased}  \\
RBAV Testbed \cite{sarantinoudis2023rosbased}  & 2023  & 1:10  & \texttimes  & \ac{ROS}  & Planning and Control  & -  \\
\citeauthor{pohlmann2022ros2based} \cite{pohlmann2022ros2based}  & 2022  & 1:10  & \texttimes  & \ac{ROS 2}, MicroROS  & Planning and Control, Communication  & -  \\
DART \cite{lyons2024dart}  & 2024  & 1:10  & \checkmark  & \ac{ROS}  & Planning and Control / System Identification  & \cite{lyons2024dart}  \\
Illi Racecar \cite{zhang2021illi}  & 2021  & 1:10  & \texttimes  & \ac{ROS 2}  & Planning and Control  & -  \\
\citeauthor{bautista-montesano2023reinforcement} \cite{bautista-montesano2023reinforcement}  & 2023  & 1:10  & \texttimes  & \ac{ROS}  & Planning and Control  & -  \\
HyphaROS Series \cite{hao-chih2017hypharos, hao-chih2018hypharosa}  & 2017  & 1:10/20  & \checkmark  & \ac{ROS}  & Planning and Control / \acs{SLAM}  & -  \\ % HyphaROS RaceCar/MiniBot/MiniBot
Cambridge Minicar \cite{hyldmar2019fleet}  & 2019  & 1:24  & \checkmark  & - (Radio)  & Planning and Control / Mixed Traffic  & \cite{mitchell2020multivehicle}  \\
\citeauthor{rupp2019fast} \cite{rupp2019fast}  & 2019  & 1:10/14  & \texttimes  & - (Wi-Fi)  & Planning and Control  & -  \\

\midrule
\multicolumn{7}{l}{\textbf{Deep Learning}}\\
\citeauthor{hong2020autonomous} \cite{hong2020autonomous}  & 2020  & -  & \texttimes  & \ac{ROS}  & Deep Learning  & \cite{asghar2023control}  \\
HydraMini \cite{wu2020hydramini}  & 2020  & -  & \texttimes  & \ac{ROS}  & Deep Learning  & -  \\
\citeauthor{caponio2024modelling} \cite{caponio2024modelling}  & 2024  & 1:10  & \texttimes  & - (Wi-Fi)  & Deep Learning  & -  \\

\midrule 
\multicolumn{7}{l}{\textbf{Aggressive Driving}}\\
F1TENTH \cite{okelly2020f1tenth}  & 2020  & 1:10  & \checkmark  & \ac{ROS}  & Autonomous Racing  & \cite{agnihotri2020teaching, xiao2021learning}  \\
BARC \cite{barcproject2024barc}  & 2016  & 1:10  & \checkmark  & \ac{ROS}  & Autonomous Racing  & \cite{gonzales2016autonomous, zhang2017autonomous}  \\
DeepRacer \cite{balaji2020deepracer}  & 2019  & 1:10  & \checkmark  & \ac{ROS}  & Autonomous Racing  & - \\
ForzaETH \cite{baumann2024forzaeth}  & 2024  & 1:10  & \checkmark  & \ac{ROS}  & Autonomous Racing  & - \\
AutoRally \cite{goldfain2019autorally}  & 2019  & 1:5  & \checkmark  & \ac{ROS}  & Agile Driving  & \cite{gandhi2021robust,  yin2023shield}  \\
Delft Scaled Vehicle \cite{baars2021control}  & 2021  & 1:10  & \texttimes  & \ac{ROS}  & Drift Control  & -  \\
AV4EV \cite{qiao2024av4ev}  & 2024  & 1:3  & \checkmark  & - (CAN bus)  & Autonomous Racing  & \cite{mangharam2024av4ev}  \\
\citeauthor{eken2020reproducible} \cite{eken2020reproducible}  & 2020  & 1:10  & \checkmark  & \ac{ROS}  & Education (Undergraduate)  & -  \\

\midrule 
\multicolumn{7}{l}{\textbf{Cloud Computing}}\\
MCCT \cite{dong2023mixed}  & 2023  & 1:14  & \texttimes  & \ac{ROS} / ZeroMQ  & Cloud Computing / Planning and Control  & \cite{yang2022multivehicle, xu2023cloudbased}  \\
HydraOne \cite{wang2019hydraone}  & 2019  & -  & \texttimes  & \ac{ROS}  & Edge Computing  & \cite{chavan2022highquality,  zeng2020fengyi}  \\
AutoE2E \cite{bai2020autoe2e}  & 2020  & 1:16  & \texttimes  & Custom (AutoE2E)  & real-time Execution  & \cite{bai2022performance}  \\
LiveMap \cite{liu2021livemap}  & 2023  & 1:18  & \texttimes  & Custom (LiveMap)  & real-time Execution / Edge Computing  & \cite{liu2023realtime}  \\
Open CLORO \cite{portaluri2019open}  & 2019  & -  & \checkmark  & - (Wi-Fi)  & Cloud Computing  & -  \\

\bottomrule
\end{tabular}
}
\end{table*}

\subsection{Classification of Small-Scale Testbeds}
\label{sec:small-scale:testbeds}

% Table
\Cref{tab:small-scale-testbeds} summarizes the small-scale testbeds we identified, including details on the testbed name, year of release, scale, whether the testbed is open-source, its middleware, focus, and related publication.  
Related publications make use of the testbed to test novel methods but do not present the testbed itself.
In the table, we group the testbeds into 9 categories, illustrated in \cref{fig:small-scale-testbeds}, to enhance readability. These categories are intended to help the reader navigate the variety of testbeds. 
Therefore, they should not be viewed as rigid classifications. For example, testbeds focusing on aggressive driving may also be relevant to the planning and control category, and testbeds designed for educational purposes could overlap with other categories depending on the educational context. 
Among the testbeds reviewed, middleware solutions vary widely. 
ROS-based middlewares dominate, with 45 out of the considered testbeds utilizing \ac{ROS}. Some testbeds (4 out of \cref{tab:small-scale-testbeds}) employ custom middleware solutions, while 20 testbeds (indicated by ``-'') do not use middleware according to our definition.

\subsection{Architectures and Middlewares of Small-Scale Testbeds}
\label{sec:small-scale:architectures}

Many small-scale testbeds use \ac{ROS} to facilitate essential tasks like communication, localization, and perception, i.e., \cite{ulhas2022chartopolis, carron2023chronos, sarantinoudis2023rosbased, paull2017duckietown}.
\Ac{ROS}-based middlewares are particularly useful for multi-agent coordination, as seen in testbeds like RoboCoPlat \cite{filho2020cooperative} and Cyclops \cite{lee2022cyclops}, where \ac{ROS} handles real-time data exchange and \ac{V2V} communication in scenarios such as platooning and sensor failure recovery. In comparison to \ac{ROS}, \ac{ROS 2} offers several advantages, particularly in terms of real-time performance, enhanced communication capabilities, and scalability, which are crucial for more complex decentralized control systems (see \cref{sec:simulation:architectures}). 
The testbeds we discovered that use \ac{ROS 2} are, among others, CPM Lab \cite{kloock_cyber-physical_2021} and Cambridge RoboMaster \cite{blumenkamp2024cambridge}.  Several testbeds, including XTENTH-CAR \cite{sivashangaran2024xtenthcar} and ForzaETH \cite{baumann2024forzaeth}, have either transitioned to \ac{ROS 2} or adopted it in parallel with their original middleware to exploit its enhanced features. Overall, although \ac{ROS} is still dominant in small-scale testbeds for \acp{CAV}, there is a clear trend towards the adoption of \ac{ROS 2}, driven by its ability to handle real-time communication and its support for decentralized control. 

Some testbeds develop custom middleware solutions to address specific requirements. Examples include:
\begin{itemize}
    \item CyPhyHouse \cite{ghosh2020cyphyhouse}, which provides a platform-independent modular middleware for distributed coordination in mobile robotic applications. It introduces a high-level programming language called Koord to simplify complex tasks similar to \ac{ROS} message handling and socket programming. CyPhyHouse uses \ac{ROS} topics during simulation but abstracts them for the developer, focusing on distributed coordination at higher level.
    \item LiveMap \cite{liu2021livemap}, which is a real-time dynamic map middleware for \acp{CAV}, focusing on automotive edge computing. It supports real-time data sharing for improved situational awareness. While it leverages \ac{ROS} for simulations, its architecture emphasizes distributed data processing and dynamic map updates, going beyond the standard \ac{ROS} frameworks.
    \item Etherware \cite{graham2009abstractions}, which is a middleware designed for networked control systems that provides a so-called virtually collocated system, simplifying the design of distributed control systems by allowing them to be developed as if running on a single machine. Etherware handles message passing, clock synchronization, and fault tolerance, offering similar benefits to \ac{ROS} but developed independently with a focus on networked control.
    \item AutoE2E \cite{bai2020autoe2e}, which is a two-tier middleware designed for real-time control in autonomous driving. It addresses CPU utilization and deadline management to ensure reliable operation under varying workloads. It is not directly based on \ac{ROS} but rather addresses limitations in standard real-time scheduling frameworks like AUTOSAR.
\end{itemize}
In summary, these custom solutions for middlewares typically build on the foundations of \ac{ROS} or similar systems, but extend them to meet the particular needs of their applications.

For testbeds that do not feature a middleware according to our definition, we provide information on the communication technologies they use. The most widely used are radio and Wi-Fi. Radio communication is used in testbeds such as Cambridge Minicar \cite{hyldmar2019fleet}, \citeauthor{la2011smallscale} \cite{la2011smallscale}, Pharos \cite{petz2012experiences}, RAVEN \cite{valenti2006indoor}, and MicroITS \cite{marchand2019microits}. Wi-Fi communication is used in testbeds like Open CLORO \cite{portaluri2019open} and \citeauthor{weinbauer2023taking} \cite{weinbauer2023taking}. Besides, \ac{DSRC} and Cellular Communication (4G/5G) are often employed in testbeds focusing on vehicular networks, such as \citeauthor{weinbauer2023taking} \cite{weinbauer2023taking}. Less commonly used communication technologies are Bluetooth and Ethernet. Among the small-scale testbeds reviewed, only ORCA \cite{liniger2015optimizationbased} and SGT \cite{bhattarai2015optimizing} use them.

The variety of testbed solutions and middleware in \ac{CAV} research reflects various requirements. 
Researchers commonly adopt \ac{ROS}-based middleware for communication and coordination, with \ac{ROS 2} increasingly chosen for enhanced performance and scalability. Customized middleware solutions address specific challenges, including distributed control and real-time map processing, while integrated physical and virtual environments support testing. 
Small-scale testbeds efficiently validate algorithms, communication protocols, and control strategies, thus bridging the gap between simulation and full-scale experiments.

\section{Full-scale Testbeds}
\label{sec:full-scale}

Full-scale vehicle testing for \acp{CAV} involves conducting experiments with real, full-sized vehicles and offers a direct, comprehensive means of assessing \acp{CAV} performance \cite{tang_survey_2023}. Such testing may involve either purpose-built platforms specifically designed for \ac{CAV} testing or on mass-produced vehicles augmented with additional sensors and computational resources \cite{karle_edgar_2024, vaio_design_2019}. By operating at realistic scales, these setups accurately capture genuine vehicle dynamics and environmental conditions.

Despite the advantages of realism, full-scale testing poses challenges related to cost, safety, and repeatability. Closed test tracks are commonly used to ensure safety \cite{jhung_end--end_2018}, but they incur considerable financial burdens from vehicle expenses, track rental, and strict safety regulations. Failures in full-scale testing also present greater risks, as even minor collisions or equipment malfunctions can cause significant damage to the vehicles and facilities.

\subsection{Classification of Full-scale Testbeds}
\label{sec:full-scale:testbeds}

\begin{figure}[tbp]
    \centering
    \includegraphics[width=\linewidth]{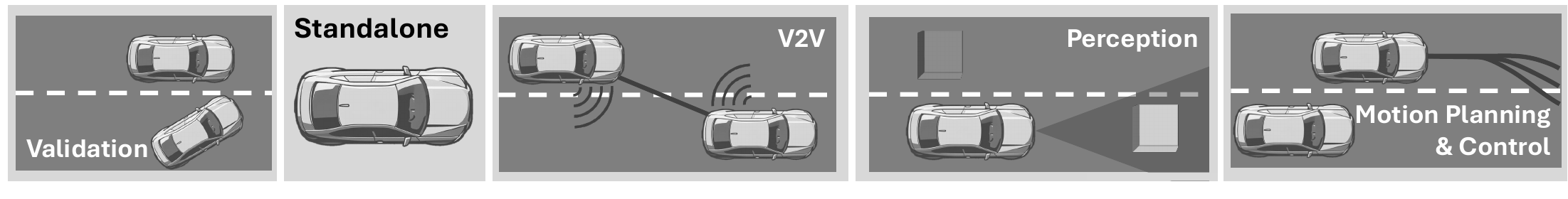}
    \caption{Illustration of the classifications for full-scale testbeds. We found five classifications: standalone, V2V communication, perception, validation, and control \& motion planning.}
    \Description{Illustration of the classifications for full-scale testbeds. We found five classifications: standalone, V2V communication, perception, validation, and control \& motion planning.}
    \label{fig:fullscale}
\end{figure}

In our literature research, we found that the development and testing of autonomous and connected vehicles requires diverse testbeds and test vehicles.
\Cref{tab:testbeds-overview} summarizes the testbeds and vehicles we identified, including details on the testbed name, year of release, scale, whether the testbed is open-source, its middleware, and related publication. 
These testbeds address various research domains such as control, perception, communication, and full-stack integration.
As noted, standalone platforms are rarely published in themselves, but more commonly as part of the evaluation for other algorithms or novel methods.
Supporting experiments in motion planning, risk assessment, and platooning protocols, we found 13 full-scale testbeds and test vehicles from the domains of motion planning, trajectory validation, V2V and control.
\begin{itemize}
\item \textbf{Standalone Testbeds:} We found multiple standalone testbeds in our research. \citeauthor{arrigoni_design_2021} \cite{arrigoni_design_2021} present a standalone \ac{CAV} research platform, which they use in a number of other publications to test motion planning concepts \cite{mentasti_beyond_2023}.
Standalone platforms, which offer end-to-end integration for autonomous driving research, enable a comprehensive approach that includes all key CAV subsystems \cite{lucchetti_toward_2023}.

\item \textbf{Motion Planning Testbeds:} \citeauthor{trauth_frenetix_2024} \cite{trauth_frenetix_2024} present the Freenetix framework for motion planning. To evaluate it, the authors used the EDGAR testbed, first presented in \cite{karle_edgar_2024}.
Other full-scale testbeds used for motion planning include \citeauthor{vaio_design_2019} \cite{vaio_design_2019}.
Additional testbeds discuss novel path planning frameworks, preview path tracking control designs, and approaches to balance safety and performance in connected truck control. 
Notable platforms in this area include a scalable platform for truck platooning \cite{alam_heavy-duty_2015}.
    
\item \textbf{Validation Testbeds:} \citeauthor{noh_risk_2017} \cite{noh_risk_2017} presented a method risk assessment model for lane changes, where they used a automated vehicle testbed in experiments to validate the concept.

\item \textbf{\ac{V2V} Testbeds:} In the area of communication, testbeds facilitate distributed interaction protocols that support information exchange across networks of \acp{CAV}. \citeauthor{di_bernardo_design_2016} \cite{di_bernardo_design_2016} propose such a protocol and evaluate it in experiments with a full-scale automated vehicle.

\item \textbf{Control Testbeds:} \citeauthor{alan_integrating_2024}  \cite{alan_integrating_2024} attempts to combine a safety-oriented controller with a performance-oriented one. The testbed is then used to demonstrate the control approach in experiments with a full-scale connected and automated truck.

\end{itemize}

\begin{table*}[tbp] 
\caption{Overview of Overview of the full-scale testbeds we found during our literature review. The testbeds are loosely grouped based on their primary focus, and they may also provide functionality beyond their primary focus.}
\centering
\label{tab:testbeds-overview}
\resizebox{\linewidth}{!}{%
\begin{tabular}{lllllll}
\toprule
\textbf{Testbed/Vehicle} & \textbf{Year} & \textbf{\#Agents} & \textbf{OSS} & \textbf{Middleware}  & \textbf{Focus} & \textbf{Related Publication} \\

\midrule
\multicolumn{6}{l}{\textbf{Motion Planning}}\\
Scania HDV \cite{alam_heavy-duty_2015} & 2015 & 1 & \texttimes & Direct-to-vehicle & Motion Planning & \cite{alam_heavy-duty_2015} \\
AstaZero Volvos \cite{vaio_design_2019} & 2019 & 1 & \texttimes & Direct-to-vehicle & Motion Planning & \cite{vaio_design_2019} \\
Platoon Control \cite{li_platoon_2020} & 2020 & 4 & \texttimes & Direct-to-vehicle & Motion Planning & \cite{li_platoon_2020} \\
EDGAR \cite{karle_edgar_2024} & 2024 & 1 & \checkmark & ROS 2 & Motion Planning & \cite{trauth_frenetix_2024} \\

\midrule
\multicolumn{6}{l}{\textbf{Vehicle Control}}\\
Yonsei Testbed \cite{jhung_end--end_2018} & 2018 & 1 & \texttimes & Direct-to-vehicle & Control & \cite{jhung_end--end_2018} \\
Hybrid MKZ \cite{xu_design_2020} & 2020 & 1 & \texttimes & Direct-to-vehicle & Control & \cite{xu_design_2020} \\
fortuna \cite{kessler_bridging_2019} & 2019 & 1 & \texttimes & MATLAB RT & Control & \cite{kessler_bridging_2019} \\
ProStar \cite{alan_integrating_2024} & 2024 & 1 & \checkmark & MATLAB RT & Control & \cite{alan_integrating_2024} \\

\midrule
\multicolumn{6}{l}{\textbf{Validation}}\\
Risk Assessment Testbed \cite{noh_risk_2017} & 2017 & 1 & \checkmark & ROS & Validation & \cite{noh_risk_2017} \\

\midrule
\multicolumn{6}{l}{\textbf{Communication}}\\
Chalmers Volvos \cite{di_bernardo_design_2016} & 2016 & 1 & \texttimes & Direct-to-vehicle & V2V MP & \cite{di_bernardo_design_2016} \\

\midrule
\multicolumn{6}{l}{\textbf{Full Platform}}\\
VNL-300 T \cite{wahba_developing_nodate} & 2018 & 1 & \texttimes & ROS & Standalone & \cite{wahba_developing_nodate} \\
BAE Wildcat Autonomous Test Vehicle \cite{kent_connected_2020} & 2020 & 1 & \texttimes & ROS & Standalone & \cite{kent_connected_2020} \\
Prototypical AV Testbed \cite{arrigoni_design_2021} & 2021 & 1 & \checkmark & ROS & Standalone & \cite{mentasti_beyond_2023} \\

\hline
\end{tabular}
}

\end{table*}

\subsection{Architectures and Middlewares of Full-scale Testbeds}
\label{sec:full-scale:architectures}

We identified three primary software architectures in the literature. These mostly fall into similar categories as the simulator-based approaches and small-scale testbeds. Either they utilize middlewares, such as \ac{ROS 2}, link directly to the vehicle or use MATLAB Simulink RT to test control approaches with strict real-time requirements.

The first and most straightforward architecture, direct-to-vehicle, used no software framework and was based on CAN communication within a monolithic application. 
This architecture typically features a single computer hardware setup, with simple and sometimes modular software architectures that connected directly to multiple sensors. 
Examples of this architecture include the testbeds by \citeauthor{xu_design_2020} \cite{xu_design_2020} and \citeauthor{vaio_design_2019} \cite{vaio_design_2019}.
Relying on a single hardware platform eliminates the challenges associated with distributed systems, as intra-process and inter-process communications are executed directly and with minimal latency. Computations within this setup are easily managed by the operating system, minimizing overhead and allowing for focus on the automated driving novelty under test \cite{vaio_design_2019}.
Direct-to-vehicle or bus-based architectures integrate a central in-vehicle computer with standard fieldbus technologies, such as CAN, enabling direct communication with embedded sensors and actuators. Execution typically relies on a standard Linux or desktop operating system, requiring no specialized deployment measures. Although these architectures are closed-source and depend on commercial licenses and compatible hardware, they are widely supported by a large user community.

Another architecture type we found utilizes ROS 2, often in a distributed system configuration composed of multiple nodes, sometimes spread across multiple computers. 
This setup is typically applied in complex, deployment-oriented environments where the system architecture must meet software requirements.
An example was presented by \citeauthor{arrigoni_design_2021} \cite{arrigoni_design_2021}, where a platform for autonomous vehicle testing was introduced. 
ROS-based systems provide the flexibility needed for more complex, modular designs and can accommodate the demands of larger-scale and distributed computing environments.

The third type of architecture, MATLAB RT, was often found for developments in the control domain and focused on real-time computation, typically operating on a single real-time machine integrated with the vehicle via CAN \cite{alan_integrating_2024}. 
A common tool was the MATLAB real-time toolkit with accompanying hardware to execute the generated code. This configuration allows for the stringent timing requirements crucial to real-time control experiments, ensuring the reliability of control loop timing. 
Although MATLAB RT is effective for experiments that require precise control loop timing, it is rarely used for full-stack implementations due to its specificity and limited scalability \cite{kessler_bridging_2019}. 
MATLAB real-time provides a proprietary framework for deploying real-time applications onto dedicated commercial hardware, commonly utilized for hardware-in-the-loop (HIL) testing. It integrates with Simulink models, using internal communication interfaces and supporting various bus systems. Real-time execution is achieved through fixed-step solvers running on a QNX Neutrino real-time operating system, coupled with corresponding real-time compilation processes. The deployment targets rapid control prototyping, ensuring a consistent fixed-step execution interval if computational resources remain sufficient. Although supported by a substantial user community, MATLAB real-time is closed-source and requires both commercial licensing and compatible hardware.

Beyond these three main architectures, we also identified two other automotive middlewares. Although our survey did not uncover publicly documented testbeds (small- or full-scale) using them, their use in industry is strongly implied by extensive surrounding work \cite{henle_architecture_2022}:

\begin{itemize}
    \item Cyber RT, developed by Apollo \cite{noauthor_apollo_nodate}, structures software as nodes, channels, tasks, and co-routines, while employing publish-subscribe or request-response concepts built on FastDDS. It triggers tasks via timers and uses callbacks for incoming messages. Suited to development environments, Cyber RT is open-source, well-documented, and can be deployed as a shared library or standalone binary.
    \item AUTOSAR (Classic and Adaptive): The Automotive Open System Architecture (AUTOSAR) is a widely adopted standard for automotive software.
    \begin{itemize}
        \item AUTOSAR Classic targets embedded, resource-constrained ECUs by defining basic software layers (e.g., microcontroller abstraction layer, ECU abstraction layer, and services layer) and providing a runtime environment where application-level software components can run deterministically. Classic AUTOSAR primarily supports statically defined communication and scheduling, making it suitable for safety-critical, real-time applications.
        \item AUTOSAR Adaptive evolves the Classic standard to address more dynamic and connected software environments. It features a service-oriented architecture, supports \ac{DDS} for communication, and provides an Execution Management component that oversees application lifecycles, concurrency, and scheduling \cite{autosar_executionManagement_2024}. Although no specific small- or full-scale testbeds using AUTOSAR Adaptive were discovered in our review, the platform offers comprehensive runtime management, safety, and security features, indicating its potential suitability for modern CAV architectures.
    \end{itemize}
\end{itemize}

\section{Requirements of Small-Scale Testbeds}
\label{sec:Requirements:small-scale}

Comparing the results from the simulations in \cref{sec:simulation} with the small-scale testbeds in \cref{sec:small-scale} reveals that far more testbeds incorporate a middleware than their simulation counterparts. This finding follows naturally from the use of real, distributed hardware in small-scale testbeds.

\begin{figure}[tbp]
    \centering
    \includegraphics[width=\linewidth]{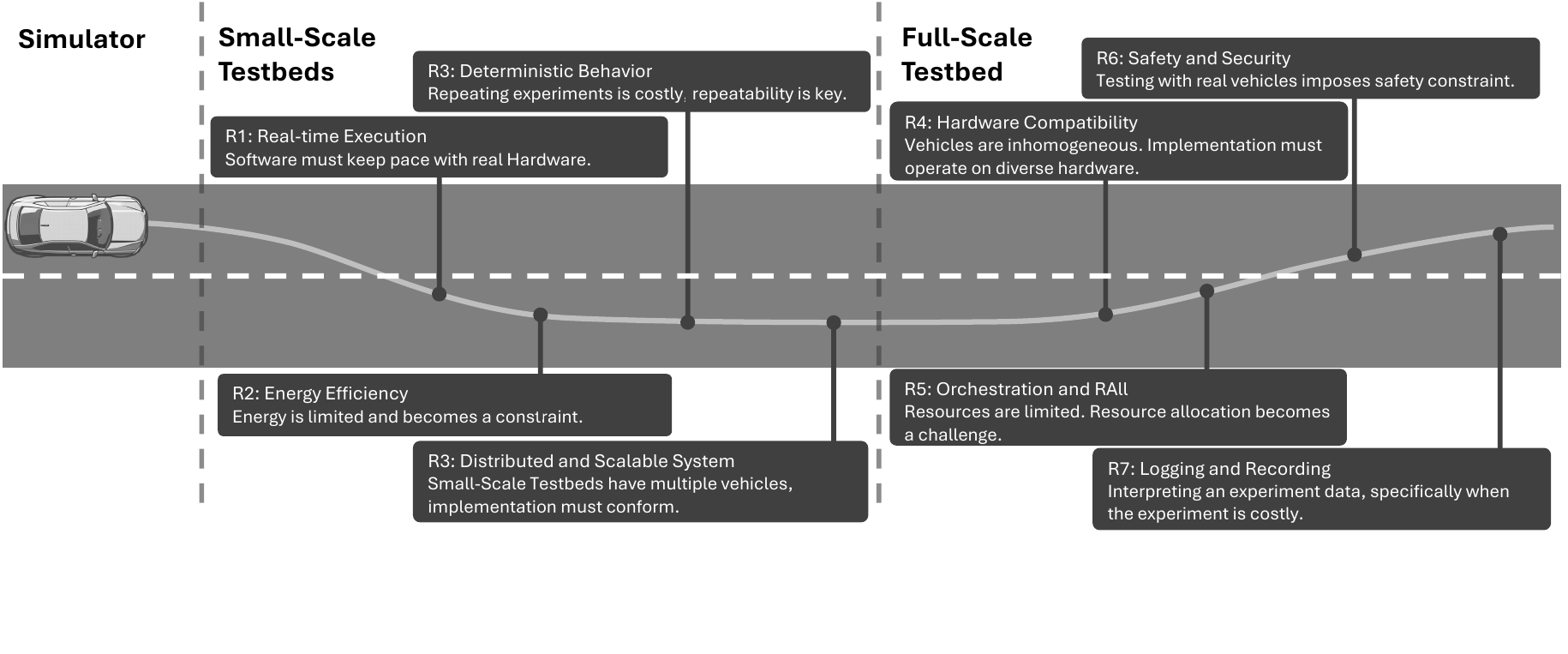}
    \caption{Illustration depicting the requirements moving from simulation to small and from small to full scale. Requirements are shown as gray boxes with a short argument for the Requirement.}
    \Description{Illustration depicting the requirements moving from simulation to small and from small to full scale. Requirements are shown as gray boxes with a short argument for the Requirement.}
    \label{fig:requirements}
\end{figure}

Based on our survey, we identified eight key challenges for the software under test, which we distilled into eight overarching requirements.
In this section, we present the first four requirements that arise specifically from the use of small-scale testbeds. The remaining four requirements, which apply primarily to full-scale testbeds, are discussed in Section \cref{sec:Requirements:full-scale}.
\Cref{fig:requirements} illustrates how moving toward increasingly realistic environments, from simulation to small-scale, and eventually to full-scale, introduces additional requirements for the software under test.

\subsection{Energy Efficiency}
\label{sec:requirements:energy}

Energy efficiency is a key priority in the transition from internal combustion to electric propulsion. To reduce both costs and environmental impact, vehicles should minimize overall energy consumption. Electric vehicles, in particular, face inherent battery constraints that result in lower ranges compared to internal combustion counterparts \cite{varga_prediction_2019}.

Although small-scale testbeds consume far less energy than their full-scale counterparts, they are similarly restricted by battery size, which limits both runtime and utilization \cite{mokhtarian2024survey}. Consequently, software supporting either small-scale or full-scale platforms must account for energy efficiency. According to \citeauthor{vahidi_energy_2018} \cite{vahidi_energy_2018}, software design can yield substantial energy savings in \acp{CAV}. However, \citeauthor{augusto_debunking_2024} \cite{augusto_debunking_2024} notes that most energy savings accumulate through drivetrain control, as the automated driving system represents, at worst, only 16\% of overall vehicle power consumption. In other words, the main potential for optimization lies in drivetrain and vehicle control, while the impact of in-vehicle software is comparatively small.

In terms of software efficiency, we identified two topics of relevance:

\begin{itemize} 
    \item \textbf{Resource Allocation:}
The middleware should enable energy-minimal resource allocation. For example, assigning compute tasks in a distributed system based on energy use can lower overall power consumption \cite{kampmann_optimization-based_2022}. Power use also decreases if \acp{ECU} can be disabled or utilized at peak efficiency.

    \item \textbf{Efficient Execution:}  
Middleware should implement methods that efficiently execute tasks. Reducing idle times and overhead can lead to further power savings \cite{faragardi_resource_2018}.
\end{itemize}

Among the reviewed middleware solutions, only AUTOSAR Adaptive explicitly addresses this concern. Its execution management component partitions resources using the operating system’s \textit{cgroups}, but its mechanisms are strictly executive, executing only a precomputed plan rather than dynamically determining resource requirements at runtime \cite{autosar_executionManagement_2024}. By contrast, \ac{ROS} and \ac{ROS 2} do not offer built-in resource allocation tools, even though \ac{ROS 2} includes a lifecycle node for managing the applications lifecycle. Approaches using DDS lack functionality to constrain or optimize software execution as well.

Beyond established frameworks, researchers have proposed additional resource-allocation strategies. \citeauthor{faragardi_resource_2018} \cite{faragardi_resource_2018} present a runtime solution based on AUTOSAR Classic, assigning tasks to processing cores in ways that minimize both core load and total active cores. Likewise, \citeauthor{kampmann_optimization-based_2022} \cite{kampmann_optimization-based_2022} propose a design-time, multi-stage optimization approach that yields power-minimal task-to-\ac{ECU} assignments, significantly reducing the total required \ac{ECU} count.

Further improvements to energy efficiency in vehicle systems can target communication in the \ac{E/E} architecture (e.g., enabling intelligent controllers to put \acp{ECU} to sleep when they are inactive \cite{schmutzler_increasing_2011}) and enhancing electric powertrain designs through refined control schemes \cite{safder_need_2024}. In summary, software-focused research focuses on optimizing resource allocation and high-intensity computation, whereas hardware research concentrates on powertrain enhancements and improved inter-\ac{ECU} communication.

\subsection{Real-Time Performance} 
\label{sec:challenges:rt-exec}

Ensuring timely software responses is critical for automated vehicles at any scale \cite{wu_oops_2021}. Control loops depend on continuous feedback, while vision-based systems such as SLAM require frequent, punctual updates \cite{betz_software_2019}. Consequently, small-scale vehicles also need prompt computations, and the underlying software must meet specific timing requirements \cite{kloock_architecture_2023}. A system with real-time guarantees certifies these timing constraints, yet most existing solutions rely on best-effort computation without explicit timing guarantees. This section examines how established middleware solutions address real-time concerns and explores additional approaches for meeting them.

Real-time performance spans both execution and communication:

\begin{itemize}
    \item \textbf{Real-Time Execution:} The middleware should enable task scheduling on each \ac{ECU} so that deadlines are met. For instance, a stability program relying on frequent wheel-speed measurements must satisfy strict timing constraints for correct operation. Such guarantees can be achieved through specialized scheduling patterns combined with real-time operating systems \cite{gemlau_system-level_2021}.

    \item \textbf{Real-Time Communication:} While scheduling on a single \ac{ECU} is necessary, acquiring required data from other \acp{ECU} in a timely manner is equally important. The middleware should therefore support meeting deadlines in networked environments. Protocols such as \ac{TSN} enable enforcement of strict communication deadline \cite{peng_survey_2023}.
\end{itemize}

Prominent middleware frameworks integrate real-time capabilities in distinct ways. For example, AUTOSAR Adaptive includes an execution management functional cluster \cite{henle_architecture_2022, autosar_executionManagement_2024}, responsible for resource allocation, execution timing control, and application lifecycle management; it depends on the operating system’s scheduler for orchestrating application execution. By supporting \ac{DDS} for communication, AUTOSAR Adaptive inherits \ac{DDS}-based real-time communication properties \cite{autosar_communicationManagement_2024}. Meanwhile, \ac{ROS 2} coordinates execution via timers and executors \cite{henle_architecture_2022, open_robotics_executors_2024}, although meeting real-time demands relies on appropriately configured \ac{ROS} middleware \cite{macenski_2022_ros2}; if a \ac{DDS}-based ROS middleware is employed, \ac{ROS 2} likewise inherits real-time communication properties. Finally, MATLAB Simulink RT, explicitly designed to guarantee real-time performance, operates on a real-time operating system to ensure feasible scheduling at acceptable CPU loads \cite{noauthor_use_nodate}, it uses a dedicated hardware solutions to enable low-latency inter-process communication.

In addition to middleware-based methods, research on real-time automotive systems offers alternative techniques.
For example, \citeauthor{becker_contention-free_2016} \cite{becker_contention-free_2016} propose a linear programming model to optimize time-triggered scheduling in classic AUTOSAR. 
Similarly, \citeauthor{biondi_achieving_2018} \cite{biondi_achieving_2018} employ the \acf{LET} paradigm to improve determinism on multicore systems, albeit with minor runtime overhead.
Virtualization is also explored through hypervisor-based multicore consolidation by the same authors.

In addition to scheduling-oriented strategies, measurement-based methods can refine timing validation. 
\citeauthor{becker_end--end_2017} \cite{becker_end--end_2017} explore end-to-end timing under varying activation patterns, and \citeauthor{kampmann_agile_2020} \cite{kampmann_agile_2020} propose agile latency estimation using testbeds. 
Similarly, \citeauthor{vilardell_cleanet_2020} \cite{vilardell_cleanet_2020} introduce CleanET to assess execution-time interference from different overlapping tasks. 

Because real-time guarantees must hold across entire systems, communication among \acp{ECU} also requires real-time approaches. 
However, middleware support may not exclusively guarantee that performance guarantees are met. 
Event-driven communication can yield unpredictable network congestion in \ac{IVN}, risking violation of soft real-time constraints.
The family of \ac{TSN} ethernet standard extensions attempts to solve this problem. 
A thorough overview of \ac{TSN} is provided by \citeauthor{ashjaei_time-sensitive_2021} \cite{ashjaei_time-sensitive_2021}, covering its development, simulation, and schedulability. 
\citeauthor{peng_survey_2023} \cite{peng_survey_2023} similarly survey the state of \ac{TSN} and its applications. 
Discussing the intersection of \ac{SDN} and \ac{TSN}, \citeauthor{hackel_software-defined_2019} \cite{hackel_software-defined_2019} propose combining \ac{SDN} with \ac{TSN} for robust network management in time-critical systems.

\subsection{Deterministic behavior}
\label{sec:challenges:rt-comm}

Ensuring that experiments are repeatable is a core part of traditional research, particularly when working with real vehicles in small-scale testbeds, where each additional run can significantly increase time and cost requirements \cite{kloock_architecture_2023, mokhtarian2024survey}. Consequently, deterministic behavior has become an important criterion in real-world testing, as deterministic execution of software frameworks ensures consistent and predictable system outcomes \cite{gemlau_system-level_2021}. When communication delays or execution orders fluctuate, even inherently deterministic software can exhibit non-deterministic behavior unless the system enforces strict determinism.

In this context, determinism denotes a system’s ability to produce predictable and consistent outcomes, executing tasks within defined time constraints and according to a fixed ordering. Many studies and programming models address ways to improve determinism in automotive systems. For example, \citeauthor{menard_achieving_2020} \cite{menard_achieving_2020} handle the intrinsic non-determinism of Adaptive AUTOSAR by integrating the \acp{LET} paradigm, adapting a traditionally event-driven mechanism into a time-deterministic execution model. Similarly, \citeauthor{gemlau_system-level_2021} \cite{gemlau_system-level_2021} examine system-level \ac{LET}, focusing on managing cause-effect chains across distributed systems and fulfilling stringent hard real-time demands.

Research on reducing non-determinism in distributed cyber-physical systems includes the work of \citeauthor{bateni_risk_2023} \cite{bateni_risk_2023}, who employ Lingua Franca as a coordination language to achieve deterministic execution in \ac{ROS 2}. Their experiments, which integrate an Autoware-based self-driving stack, illustrate the application of planned determinism in complex real-world scenarios. Additionally, \citeauthor{claraz_dynamic_2022} \cite{claraz_dynamic_2022} propose a dynamic reference architecture that consolidates multiple concepts, including \ac{LET}, to maintain predictable system behavior. Collectively, these approaches underscore the growing importance of determinism in contemporary automotive software frameworks, where repeatable and reliable testing is paramount for both academic research and industrial practice.

\subsection{Distributed Systems and Scalability}
\label{sec:Requirements:distributed}

Our survey shows that most small-scale testbeds involve multiple agents used to test automated driving functions. 
By using these small-scale platforms, researchers can benefit from dynamics that more closely resemble real vehicles while containing costs and mitigating safety risks otherwise encountered in multi-agent motion-planning research \cite{schafer_small-scale_2024}. From these operating conditions, we derive two main requirements:

\begin{itemize}
    \item \textbf{Physical Agents}: Small-scale testbeds frequently incorporate multiple agents. Tasks such as multi-agent motion planning, centralized control, or localization demand wireless communication to interact with these physical agents. Full-scale testbeds, by contrast, rarely include multiple vehicles—likely for cost considerations—and often operate as self-contained systems. Consequently, middlewares for small-scale testbeds must support reliable wireless communication across numerous agents \cite{mokhtarian2024survey}.

    \item \textbf{Distributed Systems:}
    Although small-scale testbeds inherently span multiple agents, full-scale vehicles and full-scale testbeds also feature distributed E/E architectures \cite{zhu_requirements-driven_2021}. Middlewares in these environments must provide robust communication mechanisms for \acp{IVN} \cite{karle_edgar_2024}.
\end{itemize}

In both small-scale and full-scale testbeds, the most commonly used middlewares generally support communication in distributed systems, based on established methods for device discovery via UDP multicast and IP-based protocols. For instance, \ac{ROS 2} and AUTOSAR Adaptive can each use \ac{DDS}-based communication \cite{autosar_ap_explanation_sw_arch, macenski_2022_ros2}, already implementing established features like discovery, \ac{QoS} configuration \cite{eprosima_3_2024, eprosima_5_2024}, and IP-based \ac{IVN} communication \cite{eprosima_6_2024}. Middleware frameworks with direct hardware access, by contrast, often provide only implementation-defined capabilities.

Addressing reliable communication in distributed or wireless architectures similarly rests with middleware designs. \ac{DDS}-based systems can guarantee reliable message exchange when configured with suitable \ac{QoS} policies, though these configurations may compromise timing guarantees. Achieving reliable communication remains a primary objective in \ac{V2X} research, as no standard solution currently prevails. \citeauthor{ahangar_survey_2021} \cite{ahangar_survey_2021} provide a survey on \ac{V2X} communication technologies, such as C-V2X and \ac{DSRC}, and related components within \ac{V2X} systems. Some researchers focus on dynamically assigning \ac{QoS} settings to different communication pairs, ensuring reliable communication only where necessary; best-effort protocols minimize overhead when fully reliable transmission is not required. For instance, \citeauthor{cakir_qos_2019} \cite{cakir_qos_2019} propose a negotiation-based approach for SOMEIP communications, and \citeauthor{becker_towards_2018} \cite{becker_towards_2018} configure SOMEIP \ac{QoS} to safeguard timing requirements and admission control.

\section{Requirements of Full-Scale Testbeds}
\label{sec:Requirements:full-scale}

Real vehicles pose much greater risks to the safety of other equipment, researchers conducting tests, and bystanders \cite{yurtsever_survey_2020}. 
Consequently, software has to maintain safe behavior and offer options to interrupt tests.
This increases the safety requirements for the software under test and should be considered in the software architecture \cite{nayak_automotive_2023}. 
Full-scale testbeds similarly require a requirement of how software is managed on vehicles, especially in realistic distributed systems.
Based on our survey, we summarize these transition demands into four distinct requirements and present them in this section.

\subsection{Orchestration and Resource Management}
\label{sec:Requirements:orchestration}
When testing software in full automotive systems, all participating software components must be transitioned to a ready state.
In distributed systems with multiple cooperating \acp{ECU}, such as modern vehicles, this task is non-trivial \cite{autosar_ap_explanation_sw_arch}.
Equally, the state of the software components should be tracked during vehicle operation to respond to faults or to deactivate components for resource efficiency.
To address this challenge, orchestration controls and tracks the state of software components in an automotive software system \cite{schindewolf_toward_2022}.

Future automotive software systems are likely to rely on loosely coupled services that collaborate to perform complex functionalities \cite{schindewolf_toward_2022, queiros2017sos}. 
Ensuring that these services are deployed, managed and interconnected efficiently requires robust mechanisms to handle heterogeneous hardware platforms, real-time constraints, and safety requirements \cite{sommer2025orchestrator}. 
We identify three primary aspects that middleware solutions should address to meet these needs:

\begin{itemize}
    \item \textbf{Service Orchestration:}
    Middlewares should provide an orchestration mechanism that manages automatic service deployment and lifecycle management. 
    This includes detecting and handling software failures, prioritizing safety-relevant tasks, and enabling scaling across multiple processing units \cite{schindewolf_toward_2022, sommer2025orchestrator}.

    \item \textbf{Containerization:}
    By packaging services in lightweight containers, developers can isolate dependencies, simplify deployment, and improve portability in various computing environments \cite{nayak_automotive_2023}.
    
    \item \textbf{Resource Allocation:}
    The middleware should assign computing resources to each service based on system requirements such as deadlines, fault tolerance, and real-time performance. This ensures that safety-critical tasks receive priority while optimizing overall efficiency \cite{ sommer2025orchestrator}.
    
\end{itemize}

In current middlewares, \ac{ROS 2} currently does not offer a central orchestration approach \cite{macenski_2022_ros2}. 
However, \ac{ROS 2} possesses lifecycle nodes, which allow developers to model the initial procedures, such as the activation of the sensor in the nodes.
Numerous community-developed solutions exist for \ac{ROS 2}, for example, based on Kubernetes to implement orchestration \cite{schindewolf_toward_2022, lampe2023robotkube}.
As an industrial framework, AUTOSAR Adaptive has a state management cluster \cite{autosar_ap_explanation_sw_arch}. 
This cluster is capable of transitioning function groups, groups of applications, to desired states.
Orchestration decisions are currently driven by a state machine, with provisions to respond to fault events and different driving scenarios.

In addition, several frameworks have been proposed in literature for the orchestration of services and resource management.
Most orchestration frameworks cover the distribution of services to compute hardware and their deployment.
In contrast, establishing a service composition is exclusive to the systems where the service composition can vary over time.
The orchestrator presented by \citeauthor{noguerodesign} \cite{noguerodesign} is the central component of the presented architecture. 
For example, \citeauthor{noguerodesign} \cite{noguerodesign} implements five parallel threads for task scheduling, application management, and system monitoring, while \citeauthor{nguyen_towards_2020} \cite{nguyen_towards_2020} addresses scalability via Edge-Cloud orchestration for connected and autonomous vehicles. 
Containerization supports lightweight and portable services \cite{bentaleb2022containerization}, leading to architectures that propose the use of containers in the vehicle \cite{jakobs2018dynamic} and adapt container orchestration technologies such as k3s for in-vehicle \ac{E/E} systems \cite{nayak_automotive_2023}. 
In addition, approaches promise increased resilience using dynamic orchestration \cite{schindewolf_toward_2022}. 
RobotKube \cite{lampe2023robotkube} extends these ideas to large-scale \ac{ITS} systems via Kubernetes and \ac{ROS 2}.

Several works focus on model-based design, resource management and optimization in automotive \acp{SOA}. \citeauthor{obergfell2019modelbasedd} \cite{obergfell2019modelbasedd} and \citeauthor{kugele2021modelbased} \cite{kugele2021modelbased} propose exploration of the design space for mapping hardware services, while \citeauthor{kampmann_optimization-based_2022} \cite{kampmann_optimization-based_2022} uses mathematical optimization to optimize for power consumption and latency. 
\citeauthor{mokhtarian2020dynamic} \cite{mokhtarian2020dynamic} focuses on the orchestrion of services for which \citeauthor{kampmann_optimization-based_2022}\cite{kampmann_optimization-based_2022} introduces the resource management framework.

\subsection{Security and Safety}
\label{sec:Requirements:security}
Testing in large-scale testbeds increases risk and requires additional safety considerations for the software under test \cite{alan_integrating_2024}.
Consequently, the software under test must also satisfy these functional safety requirements. 
As a fundamental component of vehicle software architecture, middleware plays an essential role in protecting communication, protecting sensitive data, ensuring fault tolerance, and complying with safety standards \cite{autosar_consortium_specification_2024}.

\begin{itemize} 
    \item \textbf{Security of Vehicle Communication:} Middleware plays a key role in ensuring communication between the various subsystems within a vehicle. Given the increasing connectivity of \acp{CAV}, middleware must implement robust security protocols, including encryption, authentication, and intrusion detection, to prevent unauthorized access that could compromise critical vehicle functions \cite{pullen_security_2024}.
    \item \textbf{Redundancy and Fault Tolerance for Safety:} Middleware ensures the reliability and safety of critical systems by supporting redundancy and fault tolerance. In case of failure of one communication channel or subsystem, middleware can switch to backup systems, ensuring that vital functions, such as steering and braking, continue to operate without interruption. This reduces the risk of accidents due to system failures \cite{autosar_consortium_specification_2024}.
\end{itemize}

Security mechanisms differ among middleware platforms. In ROS 2, for instance, \citeauthor{vilches_sros2_2022} \cite{vilches_sros2_2022} present SROS 2, a collection of developer tools and libraries designed to systematically secure computational graphs. Their approach, which follows the DevSecOps model, focuses on usability while guiding developers through creating identity and permissions certificate authorities (CAs), generating key pairs and certificates, encrypting DDS traffic via governance files, and managing permissions to protect ROS 2 applications. By contrast, AUTOSAR addresses security and safety via separate documents. The security overview \cite{autosar_security_overview_2024} outlines secure communication and data-handling measures, whereas the safety overview \cite{autosar_safety_overview_2024} discusses compliance with functional safety standards. However, detailed security guidelines for classic AUTOSAR currently appear to be unavailable.

In addition, the DDS standard, which underlies many middleware solutions, provides its own security specification \cite{omgddssecurity}. This specification defines a security model and a service plugin interface (SPI) architecture, featuring five key SPIs that support authentication, access control, cryptographic operations, logging, and data tagging. Together, these mechanisms ensure information assurance through mutual authentication, policy enforcement, auditing, and interoperability among compliant systems. Major DDS vendors, including eProsima FastDDS, Eclipse CycloneDDS, and RTI Connext, implement this specification to deliver secure DDS-based applications.

\subsection{Platform Compatibility}
\label{sec:Requirements:compatibility}
Nowadays, vehicles contain a wide range of ECUs and HPCs with varying architectures, including different processors and communication protocols.
To run the software under test in these systems, the software must interact with a number of protocols and hardware solutions \cite{zhu_requirements-driven_2021}.
Again, the middleware occupies a crucial role for this function in two main mechanisms.

\begin{itemize}
    \item \textbf{Diverse Computing Platforms:} Middleware must be able to support various processor architectures (CPUs, GPUs, FPGAs, automotive-specific processors like NVIDIA Drive, Intel's Mobileye, etc.), enabling it to operate across different hardware configurations \cite{wang_review_2024}. The middleware must operate on different operating systems and platforms, such as Linux, real-time Operating Systems (RTOS), and proprietary automotive platforms. 
    \item \textbf{In-Vehicle Communication Protocols:} The middleware must support various automotive communication standards such as Controller Area Network (CAN), FlexRay, Ethernet, and others to ensure proper data transfer between hardware components such as ECUs, sensors, and actuators \cite{zhu_requirements-driven_2021}.
\end{itemize}

Several existing middleware solutions offer varying degrees of compatibility with diverse hardware platforms and operating systems, often through out-of-the-box support or extensions to meet specialized requirements. For instance, frameworks derived from \ac{ROS 2} include mROS 2 \cite{takase2022}, which provides a minimal implementation of core \ac{ROS 2} concepts through embeddedRTPS, thereby enabling direct communication with DDS in \ac{ROS 2}. While mROS 2 supports basic publish/subscribe features and requires no dedicated agent computer, it does not currently accommodate services, actions, or parameter features, and it can be compiled for POSIX kernels, FreeRTOS, and other platforms. Another example is micro-ROS \cite{belsare2023micro}, also built upon \ac{ROS 2} but designed specifically for microcontrollers; it uses Micro XRCE-DDS (a DDS variant suited to microcontrollers) and supports FreeRTOS, Zephyr, and NuttX. Unlike mROS 2, micro-ROS integrates all core \ac{ROS 2} concepts, though it employs an agent to bridge communication between Micro XRCE-DDS and standard DDS networks.

At the same time, AUTOSAR is available in two main forms. AUTOSAR Adaptive \cite{autosar_explanationPlatformDesign_2024} targets high-performance computing systems, running on POSIX-based operating systems with a service-oriented communication model. AUTOSAR Classic \cite{autosar_classic_platform_release_2024}, in contrast, is intended for embedded hardware and resource-constrained devices, primarily using signal-based communication such as Ethernet or SOME/IP.

Several DDS implementations address compatibility with embedded systems to varying degrees.
\citeauthor{kampmann_portable_2019} \cite{kampmann_portable_2019} present embeddedRTPS, a lightweight RTPS implementation for microcontrollers, running on FreeRTOS avoiding the need for agent-based bridging and making it suitable for low-resource, standalone platform.
eProsima Micro-XRCE-DDS \cite{microxrcedds} follows a client model, where a Micro-XRCE-DDS agent bridges communication between low-resource devices and standard DDS systems. It supports FreeRTOS, Zephyr, NuttX, Linux, and Windows, with transport-layer flexibility across TCP, UDP, and serial connections.
RTI’s Connext Micro \cite{connextmicro} is a commercial DDS implementation optimized for microcontrollers, supporting 16-bit to multicore ARM and Intel CPUs. Unlike agent-based approaches, it operates independently, reducing communication overhead.
Apollo CyberRT \cite{noauthor_apollo_nodate} includes a CAN compatibility module but is limited to Intel processors and Ubuntu, restricting its use in embedded environments.
MATLAB’s Embedded Coder \cite{matlab2024embedded} enables code generation in C, C++, and CUDA for embedded applications, compliant with ISO 26262 and AUTOSAR. Simulink Real-Time \cite{matlab2024simulink} supports real-time execution on dedicated hardware.
In general, software approaches vary in embedded compatibility. Some, like embeddedRTPS and Micro-XRCE-DDS, are designed for microcontrollers, while others, such as Connext Micro, offer broader platform integration. Most support Linux and at least one RTOS, with varying degrees of transport-layer flexibility.

\subsection{Experiment Recording}
\label{sec:Requirements:logging}

In testing, logging and recording are essential for monitoring system behavior, ensuring vehicle health, and supporting diagnostic and safety investigations \cite{autosar_ap_explanation_sw_arch, bickelhaupt_towards_2024, stoffel_distributed_2023}. 
We identified four ways in which middleware contributes to logging and recording within vehicle systems.

\begin{itemize}
    \item \textbf{Troubleshooting and Diagnostics:} Middleware is responsible for handling communication between various vehicle subsystems, such as sensors, controllers, and actuators. Logging helps track interactions and events, providing data when diagnosing issues or identifying system failures. This can be helpful in understanding the root cause of a malfunction, be it hardware, software, or network-related \cite{autosar_ap_explanation_sw_arch, bickelhaupt_towards_2024}.
    \item \textbf{Vehicle Health Management:} Middleware in vehicles can support continuous health monitoring of critical systems, such as engine, brake or powertrain. By logging data such as sensor readings, error codes, and system states, you can predict maintenance needs and trigger alerts before failures occur. This can help prevent breakdowns and ensure that the vehicle remains operational and safe \cite{autosar_ap_explanation_sw_arch}.
\end{itemize}

Most existing middleware solutions provide logging and recording features that cater to a variety of use cases, whether through standard toolsets or more customizable options. For instance, many DDS-based platforms come with integrated capabilities: FastDDS offers \textit{Record and Replay}, while RTI Connext provides \textit{Connext Logging} and the \textit{RTI Recording Service} for message capture. In ROS 2, developers can use ROS Bags to record and replay topic data, collect runtime logs via the \textit{/rosout} interface, employ the \textit{rcl\_logging} framework for structured logging, and analyze performance through \textit{ROS 2\_tracing} \cite{bedard_ros2_tracing_2022}. AUTOSAR employs the Diagnostic Log and Trace (DLT) protocol \cite{autosar_log_and_trace_protocol_2024}, originally introduced for the Classic platform and extended into the Adaptive platform to log diagnostic data and trace ECU activities. Meanwhile, MATLAB Simulink offers logging and profiling blocks for signal data capture, diagnostics, and events, and Apollo CyberRT includes its own record reader/writer, analogous to ROS Bag files \cite{noauthor_apollo_nodate}.

\section{Transitioning Capabilities}
\label{sec:transitioning}

In this section, we address the core question of how to evaluate new \ac{CAV} perception, planning and control approaches.
To prove whether a new approach is meaningful, the evaluation must be convincing and scientifically rigorous.
In the literature, consensus and best practices have emerged in the evaluation of these approaches.
Based on our findings in this survey, in \cref{sec:transitioning:sim-full} we provide recommendations to transition testing from scale to scale.
In addition, we summarize these recommendations in five findings in \cref{sec:transitioning:use-case-drive}. 
Testing in virtual experiments, at small scale and finally using full-scale testbeds introduces progressively stricter requirements for the experiment, vehicle, and software architecture.
Although \simulators\ allow for simplified and consequence-free testing, incorporating physical hardware demands real-time constraints, deterministic execution, and robust communication protocols to account for real-world complexities.
For this reason, we summarize our requirements and discuss the transitions required first from simulation to scaled testing and second from small-scale to full-scale testing.

\subsection{From Simulation to Small-Scale to Full-Scale}
\label{sec:transitioning:sim-full}

In this section, we found three key requirements when moving from simulation to small-scale experiments.
In simulation frameworks, such as CARLA and Gazebo, the execution time can be paused or accelerated to accommodate computational delays. 
However, small-scale platforms, such as F1TENTH\cite{okelly2020f1tenth}, Duckietown\cite{paull2017duckietown} and CPM-lab \cite{kloock_cyber-physical_2021} require bounded-time sensor inputs and control outputs for correct algorithm function.
Hence, implementations must meet real-time constraints using tools such as MATLAB RT or custom frameworks.
Physical experiments introduce non-deterministic factors such as battery fluctuations, CPU load, and packet drops \cite{kloock_cyber-physical_2021}, making repeatability challenging. Consequently, deterministic software architectures, specialized frameworks \cite{kloock_architecture_2023}, or lingua franca\cite{lee_time_2021} are necessary to ensure consistent results and meaningful statistics.
Virtual experiments often abstract network constraints. Small-scale multi-agent testbeds, such as IDS3C\cite{stager2018scaled} and MiniCity\cite{buckman2022evaluating} rely on robust wireless communication, which may suffer from packet loss and delays. 
Accordingly, \ac{QoS} and distributed system requirements must be incorporated to manage these imperfections.

When transitioning from small-scale environments to full-scale vehicles, the following requirements become critical.
Safety and security are most important in full-scale vehicles \cite{pullen_security_2024}. 
New methods must be integrated into existing safety systems, and only mature algorithms with low failure probabilities should be tested on full-scale vehicles. Common solutions include reusing parts of the \ac{OEM} architecture or equipping racing vehicles with emergency hardware-stops, such as in the Indy Autonomous Challenge \cite{IndyAutonomousChallenge}.
Simulations often assume abundant computing resources, while real vehicles have limited onboard capabilities \cite{nayak_automotive_2023}. 
Thus, resource allocation and lifecycle orchestration become essential. Software architectures must manage startup, operation, and shutdown procedures within constrained environments.
Because full-scale testing is costly, rerunning experiments may not be feasible. Consequently, complete data collection is important \cite{kloock_architecture_2023}. 
Systems should record sufficient information to permit after-the-fact analysis without repeating hazardous or expensive trials.
Full-scale vehicles employ complex E/E architectures with multiple controllers, buses, and HPCs \cite{zhu_requirements-driven_2021}. 
Implementations must ensure compatibility with hardware and software, respecting automotive integration constraints.

\subsection{But what Testbed to Choose?}
\label{sec:transitioning:use-case-drive}
In this section, we examine different use cases commonly encountered in a \ac{CAV} research and how to evaluate them. 
To prove conclusively whether a novel approach is worthwhile, the evaluation must be convincing and scientifically rigorous.
Based on our findings in this survey, we provide recommendations for evaluation. 
Specifically, we suggest the kind of testbed to use and the \ac{E/E} and software architecture. 
We have summarized these recommendations in 4 findings.

\begin{finding}{Testing real-time execution is best done using dedicated real-time hardware and software.}{}
    Small-scale prototypes help assess control-loop timing constraints. 
    Full-scale systems must meet strict safety and latency requirements. 
    Architectures range from custom real-time hardware and software to ROS with real-time patches. 
    We recommend hardware-in-the-loop tests for latency analysis, followed by scaled or full-scale validation.
\end{finding}

\begin{finding}{Testing planning and control, racing, and platooning are best done on a small-scale testbed.}{}
    Small-scale testbeds allow for fast algorithm development and iteration while maintaining reduced risk. 
    For novel planning or platooning approaches, small-scale testing is preferable, as the vehicle behavior and distributed nature of the system is highly realistic while not incurring the risk of full-scale testing.
    Full-scale testbeds provide realistic real-world validation for mature algorithms.
    Architectures for small-scale testbeds typically use ROS or \ac{ROS 2}, while direct CAN connections are also common on full scale. 
    We recommend starting with \simulators\ such as Gazebo or CARLA, then testing on small-scale testbeds before transitioning to full-scale validation.
\end{finding}

\begin{finding}{Traffic management testing should be performed in simulation.}{}
    City-scale \simulators\ are essential for modeling traffic flows and evaluating new policies, with tools like SUMO providing a scalable foundation. 
    Small-scale testbeds replicate intersections and corridors for physical validation, while large-scale facilities enable full-scale testing with real V2X components. 
    Due to the number of agents, traffic or high agent count simulators commonly directly interact with the agent software for performance. When testing intersections or sections of traffic behavior, small-scale testbeds often make use of ROS or \ac{ROS 2}. 
    We recommend testing traffic behavior and flow in simulation and small scenarios using small-scale testbeds.
\end{finding}

\begin{finding}{Edge computing and V2X testing are commonly performed in small or full-scale testbeds.}{}
    Small-scale testbeds provide a platform for evaluating off-loaded sensor processing, distributed control, and, to some extent, communication protocols. 
    These setups commonly use ROS for connectivity, enabling early-stage testing of hybrid architectures and network reliability. 
    Full-scale systems should be used to realistically evaluate V2X and V2V use cases, as simulation of DSRC or C-V2X complicates simulation.
    We recommend starting with network \simulators\ and small-scale proof-of-concept testbeds before incrementally scaling to full-scale on-road systems for comprehensive validation.
\end{finding}

\section{Conclusion}
\label{sec:conclusion}

In this survey, we investigated the relationship among software architecture, methods under test, and \ac{CAV} testbeds. We found that novel \ac{CAV} approaches, spanning a wide range of domains, are typically assessed through three primary methods: simulation experiments, small-scale testbeds, and full-scale testbeds. After presenting 30 simulators, 54 small-scale testbeds, and 13 full-scale testbeds, along with the software architectures that enable these environments, our analysis suggests that the choice of both architecture and testbed scale is largely driven by simplicity and convenience.

Based on our findings, we compiled best practices for selecting testbed scale and supporting software. Motion-planning and control algorithms benefit most from small-scale platforms, while traffic management systems are most effectively evaluated in simulation. Perception tasks can be validated at either small or full scale, and \ac{V2X} experiments are best suited to full-scale testing. In addition, we identified key requirements that arise when transitioning from simulation to small-scale settings, emphasizing the importance of performance, timing constraints, and network connectivity in large-scale evaluations.

\begin{acks}
This research is accomplished within the project ”AUTOtechagil” (FKZ 01IS22088A). We acknowledge the financial support for the project by the Federal Ministry of Education and Research of Germany (BMBF). \\
This research is accomplished within the project ”Harmonizing Mobility” (19FS2035A). We acknowledge the financial support for the project by the German Federal Ministry for Digital and Transport (BMDV). \\
This research is supported by the Deutsche Forschungsgemeinschaft (German Research Foundation) with the grant number 468483200. \\
This research is supported by the Collaborative Research Center / Transregio 339 of the Deutsche Forschungsgemeinschaft (German Research Foundation).
\end{acks}

\acrodefplural{OS}[OS's]{Operating Systems}

\begin{acronym}[aaaaaaaaaa]\itemsep0pt
    % A
    \acro{AA}{Adaptive Application}
    \acro{AD}{Automated Driving}
    \acro{AE}{Automotive Ethernet}
    \acro{AGV}{Automated Guided Vehicle}
    \acro{AP}{Adaptive Platform}
    \acro{API}{Application Programming Interface}
    \acro{ARA}{AUTOSAR Runtime for Adaptive Applications}
    \acro{ASOA}{Automotive Service-Oriented Software Architecture}
    \acro{AUTOCONT}{Automotive Container}
    \acro{AUTOSAR}{AUTomotive Open System ARchitecture}
    \acro{AV}{Automated Vehicle}
    % B
    % C
    \acro{CAN}{Controller Area Network}
    \acro{CAN-FD}{Controller Area Network and Flexible Data-Rate}
    \acro{CAV}{Connected and Automated Vehicle}
    \acro{C-ITS}{Cooperative Intelligent Transportation System}
    \acro{CM}{Communication Management}
    \acro{COM}{Communication}
    \acro{COTS}{Commercial-off-the-shelf}
    \acro{CPEP}{Cyber-Physical Event Processing}
    \acro{CPS}{Cyber-Physical System}
    \acro{CSA}{Cooperative Swapping Approach}
    \acro{CTA}{Cooperative Transform Approach}
    % D
    \acro{DAC}{Digital-to-Analog Converter}
    \acro{DCU}{Domain Control Unit}
    \acro{DDS}{Data Distribution Service}
    \acro{DEAR}{Discrete Events for \ac{AUTOSAR}}
    \acro{DL}{Deep Learning}
    \acro{DSRC}{Dedicated Short-Range Communications}
    % E
    \acro{ECU}{Electronic Control Unit}
    \acro{E/E}{Electrical/Electronic}
    \acro{EM}{Execution Management}
    % F
    \acro{FBP}{Fractional-type Basic Period}
    \acro{FC}{Function Cluster}
    \acro{FPGA}{Field Programmable Gate Array}
    % G
    \acro{GPU}{Graphics Processing Unit}
    % H
    \acro{HMI}{Human Machine Interface}
    \acro{HPC}{High-Performance Computer}
    % I
    \acro{IDS}{Intrusion Detection System}
    \acro{IoT}{Internet of Things}
    \acro{IP}{Internet Protocol}
    \acro{IPC}{Inter-Process Communication}
    \acro{IVN}{In-Vehicular Network}
    \acro{ITS}{Intelligent Transportation System}
    % J
    % K
    % L
    \acro{LET}{Logical Execution Time}
    \acro{LIN}{Local Interconnect Network}
    % M
    \acro{MCU}{Microcontroller Unit}
    \acro{ML}{Machine Learning}
    \acro{MOST}{Media-Oriented Systems Transport}
    \acro{MQTT}{Message Queue Telemetry Transport}
    % \acro{MQTT}{Message QTT}
    % N
    \acro{NDN}{Named Data Networking}  
    % O
    \acro{OEM}{Original Equipment Manufacturer}
    \acro{OMG}{Object Management Group}
    \acro{OS}{Operating System}
    \acro{OTA}{Over-the-Air}
    % P
    % \acro{POSIX}{Portable Operating System Interface}
    % Q
    \acro{QoS}{Quality of Service}
    % R
    \acro{ROS}{Robotic Operating System}
    \acro{ROS}{Robotic Operating System 1}
    \acro{ROS 2}{Robotic Operating System 2}
    \acro{RTOS}{real-time Operating System}
    \acro{RTPS}{real-time Publish-Subscribe}
    % S
    \acro{S2S}{Service-to-Signal}
    \acro{SDV}{Software-Defined Vehicle}
    \acro{SDN}{Software-Defined Network}
    \acro{SL}{System-Level}
    \acro{SLAM}{Simultaneous Localization and Mapping}
    \acro{SM}{State Management}
    \acro{SOA}{Service-oriented Architecture}
    \acro{SoC}{System-on-a-Chip}
    % T
    \acro{TCP}{Transmission Control Protocol}
    \acro{TSN}{Time-sensitive Networking}
    % U
    \acro{UCM}{Update and Configuration Management}
    \acro{UDP}{User Datagram Protocol}
    \acro{ULL}{ultra-low latency}
    % V
    \acro{V2V}{Vehicle-to-Vehicle}
    \acro{V2X}{Vehicle-to-Everything}
    \acro{VANET}{Vehicular Ad-Hoc Network}
    \acro{VM}{Virtual Machine}
    \acro{VSS}{Vehicle Signal Specification}
    % W
    % X
    % Y
    % Z
    \acro{ZCU}{Zone Control Unit}
\end{acronym}

\bibliographystyle{ACM-Reference-Format}
\bibliography{literature.bib}

\appendix

\end{document}